\documentclass{aastex631}
\usepackage{float}
\usepackage{placeins} 
\usepackage{amsmath}
\begin{document}

\title{Stellar Ages: A Code to Infer Properties of Stellar Populations}

\correspondingauthor{Joseph J. Guzman}
\email{jjguzman@fsu.edu}

\correspondingauthor{Jeremiah W. Murphy}
\email{jwmurphy@fsu.edu}

\correspondingauthor{Andr\'es F. Barrientos}
\email{abarrientos@fsu.edu}

\author[0000-0001-8878-4994]{Joseph J. Guzman}
\affiliation{Department of Physics, Florida State University, 77 Chieftan Way, Tallahassee, FL 32306, USA}

\author[0000-0003-1599-5656]{Jeremiah W. Murphy}
\affiliation{Department of Physics, Florida State University, 77 Chieftan Way, Tallahassee, FL 32306, USA}

\author[0000-0001-8196-7229]{Andr\'es F. Barrientos}
\affiliation{Department of Statistics, Florida State University, 117 N. Woodward Ave, Tallahassee, FL 32306, USA}

\author[0000-0002-7502-0597]{Benjamin F.~Williams}
\affiliation{Astronomy Department, University of Washington, Box 351580, U.W. Seattle, WA 98195-1580, USA}

\author[0000-0002-1264-2006]{Julianne J.~Dalcanton}
\affiliation{Astronomy Department, University of Washington, Box 351580, U.W. Seattle, WA 98195-1580, USA}
\affiliation{Center for Computational Astrophysics,
Flatiron Institute, 162 Fifth Avenue, New York, NY 10010, USA}

\begin{abstract}

We present a novel statistical algorithm, \textit{Stellar Ages}, which currently infers the age, metallicity, and extinction posterior distributions of stellar populations from their magnitudes. While this paper focuses on these parameters, the framework is readily adaptable to include additional properties, such as rotation, in future work. Historical age-dating techniques either model individual stars or populations of stars, often sacrificing population context or precision for individual estimates. Stellar Ages does both, combining the strengths of these approaches to provide precise individual ages for stars while leveraging population-level constraints.  We verify the algorithm's capabilities by determining the age of synthetic stellar populations and actual stellar populations surrounding a nearby supernova, SN 2004dj.  In addition to inferring an age, we infer a progenitor mass consistent with direct observations of the precursor star. The median age inferred from the brightest nearby stars is $\log_{10}$(Age/yr) = $7.19^{+0.10}_{-0.13}$, and its corresponding progenitor mass is $13.95^{+3.33}_{-1.96}$ $\text{M}_{\odot}$.

\end{abstract}

\keywords{Supernovae (1668), Core-collapse supernovae (304), Massive stars (732), Stellar populations (1622)}

\section{Introduction} \label{sec:intro}
Determining the ages of stars plays a pivotal role in understanding stellar evolution, galactic formation, and the chemical enrichment history of the universe. Accurately dating stars reveals their life cycles, from formation in molecular clouds, to their ultimate fates as white dwarfs, neutron stars or black holes.
Age distributions reveal a system's star formation history (SFH) \citep{Tinsley1980_StellarEvolution, Harris_2009_SFHofLMC}, chemical enrichment \citep{feltzing2013_MW_AgeMet}, and dynamical evolution \citep{Eggen1962_MW_AgeDynamics, Sellwood_Binney_2002_RadialMixing, Vesperini_2013_DynamicsinGC}. These distributions trace phases of galaxy formation \citep{Rezini2006_StarDiagnosticsforGalaxies}, element production \citep{Thomas_2005}, and indicate progenitor masses for events like supernovae \citep{Smartt2009_Constraintson2p}. For instance, stellar ages offer insight into galaxy formation phases and the production of key elements such as iron and alpha-elements. Studies on the Milky Way’s thick disk, such as \citet{Bergemann2014_GAIA_MW}, highlight how chemical evolution is closely connected to stellar ages, helping us understand element build-up over billions of years. In the study of stellar clusters, precise age measurements have shown how local environment and initial cluster mass affect star formation rates and cluster evolution \citep{Baade1944_ClusterAges, DeAngeli_2005_GlobularClusterAges, Marín-Franch_2009_ClusterAgeDispersion}.

Beyond tracing chemical enrichment and galactic dynamics, stellar ages also play a critical role in estimating the progenitor masses of core-collapse supernovae (CCSNe). Stellar evolution theory predicts that massive stars, with masses $\gtrsim$ 8 M$_{\odot}$, undergo core collapse \citep{woosley2002}. Yet determining which stars explode and leave behind neutron stars (NS) versus those that collapse to form black holes remains uncertain. Three broad strategies characterize the progenitors of CCSNe: (1) modeling the brightness evolution of supernovae to estimate ejecta mass or the size of the exploding iron core \citep{bartunov1994,moriya2011,kasen2009,hillier2012,morozova2015,barker2023}. (2) serendipitous pre-imaging to analyze the progenitor's brightness and color for mass estimation \citep{smartt_2015, VanDyk2017, strotjohann2023}, and (3) age-dating the surrounding stellar population to estimate the zero-age main sequence (ZAMS) mass of the exploded star \citep{VanDyk_1999, Gogarten_2009, Maund2017, Williams_2018, Koplitz_2023, diaz-rodriguez2021}. The third strategy involves modeling the ages of stars within $\sim$100 pc of the supernova to construct a color-magnitude diagram (CMD) that reflects populations age, to infer the progenitor's ZAMS mass. Age-dating has become instrumental in deriving progenitor masses for hundreds of CCSNe. This is particularly important when direct observations of the supernova or progenitor are unavailable. For instance, \citet{Williams_2018} derived progenitor masses for historical SNe within $\sim$8 Mpc, and \citet{jennings2014} determined the ages and progenitor masses for 115 SNRs. Based on this sample, \citet{diaz-rodriguez2018} estimated the progenitor mass distribution, identifying a minimum CCSN mass of 7.3$\pm 0.1$ M$_{\odot}$, the most precise statistical estimate to date.

The age distribution of a stellar population is critical for understanding time-delay phenomena in astrophysics. Delay-time distributions (DTDs) describe the temporal lag between the formation of a stellar population and the occurrence of events like CCSNe or Type Ia supernovae \citep{Totani2008_DTD, Maoz2012_DTD, Rodney_2014DTD}. For instance, DTDs for CCSNe are closely tied to the lifetimes of massive stars, with progenitors having initial masses $\gtrsim$ 8 M$_\odot$ exploding within $\sim$50 Myr \citep{Smartt2009_Constraintson2p, Zapartas2019_DTD}. Accurate stellar age dating thus provides a direct means of constraining DTDs and by extension, the progenitor mass function. \textit{Stellar Ages} offers a means of linking observed stellar populations to their supernova progenitors and corresponding delay times.

This need for precise age estimates motivated the development of {\it Stellar Ages}, and the remainder of this introduction explores the insights and challenges in determining the ages and masses of CCSN progenitors. Determining stellar ages is a complex task due to mild degeneracies among age, metallicity, extinction, and uncertainties in stellar models. An early approach involved fitting isochrones to individual stars' magnitudes and colors \citep{Worthey1994, Pont2004_iso}. This straightforward method assigns an age by identifying the isochrone that passes closest to the star's observed CMD position, often incorporating uncertainties by sampling around photometric errors. However, this approach treats each star in isolation, ignoring the broader context of the stellar population, thus missing key constraints.

To address these limitations, methods such as synthetic CMD fitting have been developed, allowing researchers to model entire stellar populations by comparing observed CMDs with theoretical predictions based on the Initial Mass Function (IMF), binary fraction and SFH \citep{dolphin2002, dolphin2013, Aparicio_2009_SynthCMD, Gallart2005_SynthCMD}.
Fitting color-magnitude distributions of samples of stars has become the standard technique over the past few decades. for determining their age distribution.  For example, the software package MATCH infers stellar ages or SFHs, by modeling star counts in CMDs \citep{dolphin2002, dolphin2013}. MATCH uses Poisson statistics to estimate the likelihood of observed star counts in individual CMD bins. While effective for reconstructing SFHs over large regions, this approach has limitations. Binning leads to information loss, and bins with sparse populations introduce significant uncertainties, particularly for evolved stars. Additionally, since the majority of stars reside on the Main Sequence (MS), the algorithm's reliance on low-number statistics for evolved, massive stars limits its reliability when dating individual bright stars.

While MATCH has been a prominent tool for CMD fitting, many other codes have also contributed to stellar age and SFH determination. To highlight a few, StarFISH \citep{Harris_2001_StarFISH} implements a Bayesian approach to model CMDs using synthetic stellar populations, emphasizing the likelihood of observed star counts. Another widely used tool, PARAM \citep{daSilva2006_PARAM, Rodrigues2014_PARAM, Rodrigues2017_PARAM}, estimates stellar parameters (including age) by comparing observational data to precomputed isochrone grids. Additionally, ASteCA \citep{Perren2015_Asteca} combines statistical tests and Bayesian inference to analyze star cluster properties, including age distributions. Each of these tools has strengths in specific contexts, yet they often face similar challenges, including reliance on binned CMD data or assumptions about single-age populations.

Here, we present a new statistical technique that combines the strengths of fitting isochrones to individual stars and the synthetic CMD approach. \textit{Stellar Ages} accomplishes this by obtaining precise age estimates for individual stars, while leveraging the full population-level constraints offered by synthetic CMDs through the use of joint probability density functions \citep{Gelman2013_BayseianAnalysis, Bishop2006_PatternRecog}. By incorporating Bayesian inference and probabilistic modeling, {\it Stellar Ages} delivers robust stellar age estimates that are critical for environments with mixed populations, such as supernova progenitor sites. For example, traditional methods might yield large uncertainties for the brightest evolved stars, but \textit{Stellar Ages} poses the unique statistical question `What is the most likely age of this observed star?' For sufficiently bright stars, this yields more precise results. For example, a star with an absolute magnitude of -8.6 in F814W is probably younger than 10 Myr, with a minimum mass of about 19 M$_\odot$. Thus, even with only a few bright stars, we can constrain the age of the population with high certainty. By shifting the statistical question from `How many stars do we expect for a given age?' to `What is the likely age of each star?', we better utilize information from the brightest stars.

\textit{Stellar Ages} is adaptable to data from a variety of astronomical instruments, including the Hubble Space Telescope (HST) and GAIA.  It is also straight forward to incorporate more than two wavelength bands.  The algorithm has already been applied using GAIA and Hipparcos data in \citet{murphy2024}, and we demonstrate its application with HST data in this paper. Its versatility across observatories ensures that \textit{Stellar Ages} can provide new insights into stellar populations in both nearby clusters and distant galaxies.

The structure of this paper is as follows: Section 2 describes the \textit{Stellar Ages} algorithm and its statistical framework. Section 3 presents verification tests using synthetic and real data, focusing on results and implications. Section 4 discusses the broader impact of our findings and outlines potential future research. Finally, Section 5 concludes with a summary of our key contributions and suggests avenues for further study.

\section{Technique} \label{sec:technique}
\textit{Stellar Ages} infers the posterior distributions for age ($t$), metallicity ([M/H]) and median extinction, $\tilde{A}_{\text{V}}$. $\theta = \{t, \text{[M/H]}, \tilde{A}_{\text{V}}\}$.  Schematically, the posterior distribution is:
\setlength{\abovedisplayskip}{5pt}
\setlength{\belowdisplayskip}{5pt}
\begin{equation}
P(\theta | D) \propto \mathcal{L}(D | \theta) P(t) P(\rm{[M/H]}) P(\tilde{A}_{\text{V}}) \, .
\end{equation}
The data ($D$) are the set of magnitudes for all stars.  The likelihood $\mathcal{L}(D| \theta)$ models the observed brightnesses of each star after convolving with the initial mass function, dust extinction distribution, and observational uncertainties. The predicted magnitude of a star in given band ($a$) is:
\setlength{\abovedisplayskip}{5pt}
\begin{equation}
    m_{a} = \tilde{m}_{a}(M, t, [\text{M}/\text{H}]) + f(a)A_{\text{V}} + e_{a}.
    \label{eqn:MagA}
\end{equation}
Where: $m_{a}$ is the observed magnitude in wavelength band ($a$). $\tilde{m}_{a}$ is the model-predicted magnitude based stellar evolution models (as a function of initial mass $M$, age $t$, and metallicity [M/H]). $A_{a} = f(a)A_{\text{V}}$, where $A_{\text{V}}$ represents the extinction parameter in the Johnson-Cousins band $\text{V}$ and $f(a)$ represents the extinction scaling factor from $\text{V}$ to $a$. Finally, $e_{a}$ represents random observational error in magnitude band $a$. The model for $b$ follows a similar format to that for $a$. 

In this manuscript, we present data using the stellar evolution model PARSEC v1.2s \citep{Bressan2012, Chen2014, Chen2015, Fu2018}, however it should be emphasized that any stellar evolution model which yields predicted magnitudes is applicable in \textit{Stellar Ages} for inferences. For instance, \textit{Stellar Ages} is already compatible with the Mesa Isochrones and Stellar Tracks (MIST) \citep{MIST2016_0, MIST2016_1} and PARSEC 2.0 \citep{Costa2019a_Parsec2.0, Costa2019b_Parsec2.0, Nguyen2022_Parsec2.0} as demonstrated in \citet{murphy2024}.

The joint likelihood of observing a star with magnitudes $m_{a}$ and $m_{b}$ is:
\begin{equation}
\begin{split}
\mathcal{L}(m_{a}, m_{b}| \theta) = & \int p(m_{a}|\tilde{m}_{a}, A_{\text{V}}) \, p(m_{b}|\tilde{m}_{b}, A_{\text{V}}) \, p(\tilde{m}_{a}|t, [\text{M}/\text{H}], M) \, p(\tilde{m}_{b}|t, [\text{M}/\text{H}], M) \\
& \times p(A_{\text{V}}|\tilde{A}_{\text{V}},\sigma_{\text{V}}) \, p(M) \, dM \, dA_{\text{V}} \, d\tilde{m}_{a} \, d\tilde{m}_{b}
\label{eq:likelihood}
\end{split}
\end{equation}

\begin{figure}[H]
    \centering
    \begin{minipage}[b]{0.45\textwidth}
        \centering
        \includegraphics[width=\textwidth]{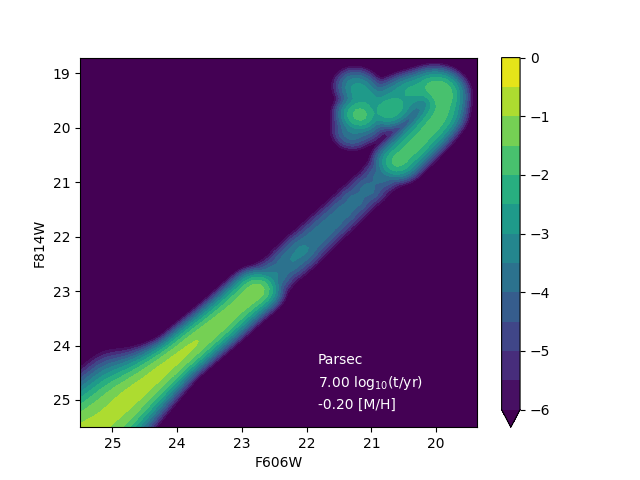}
    \end{minipage}
    \hfill
    \begin{minipage}[b]{0.45\textwidth}
        \centering
        \includegraphics[width=\textwidth]{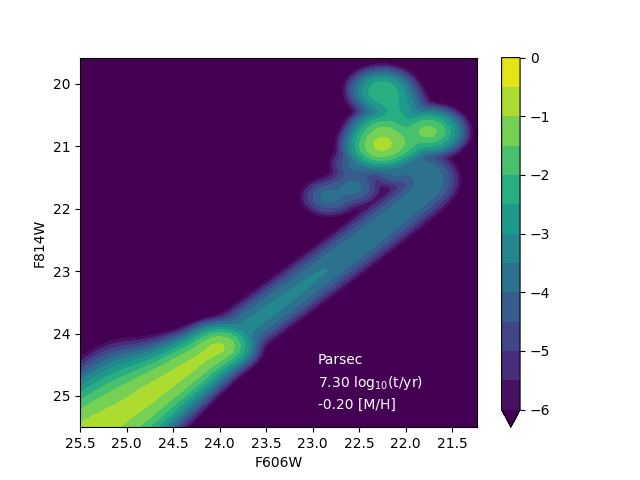}
    \end{minipage}
    \caption{\small Joint probability density functions for two magnitudes, given an age and metallicity.  These two examples have a metallicity of $[M/H] = -0.20$, $\tilde{A}_{\text{V}} = 0.0$, and ages of $\log_{10}$(t/yr) = $7.00$ and $7.20$. These density functions represent the likelihoods of observing a star at the F814W and F606W bands of the ACS HRC imager. These joint probability density functions are a convolution of measurement uncertainties with stellar evolution models, initial mass function, and an extinction distribution.}
    \label{fig:Likelitables}
\end{figure}

This likelihood represents the joint probability density function for a given age, metallicity and median extinction. Where $p(m_{a}|\tilde{m}_{a},A_{\text{V}})$ and $p(m_{b}|\tilde{m}_{b},A_{\text{V}})$ are Gaussian functions whose widths are given by observational uncertainties. Whereas $p(\tilde{m}_{a}|t, [\text{M}/\text{H}], M)$ and $p(\tilde{m}_{B}|t, [\text{M}/\text{H}], M)$ are Dirac-delta functions from the stellar evolution models. $p(A_{\text{V}}|\tilde{A}_{\text{V}},\sigma_{\text{V}})$ follows a log-normal distribution following \citep{Dalcanton_2015}, where $\tilde{A}_{\text{V}}$ represents the median extinction and $\sigma_{\text{V}}$ is a dimensionless parameter that sets the width and skewness of the log-normal distribution. For this study, we adopt a prior of $\sigma_{\text{V}}$ = 0.24, derived by \citet{Dalcanton_2015}. In that work, they initially set a prior of $\sigma_{\text{V}}$ = 0.3 based on Milky Way data \citep{Kainulainen2009_MWExtinction}, and later refined it to $\sigma_{\text{V}}$ = 0.24 for M31 by comparing Red Giant Branch stars in the foreground and background. However, this value remains largely unconstrained for many extragalactic environments due to incompleteness of dust maps.
\begin{equation}
p(A_{\text{V}}|\tilde{A}_{\text{V}}, \sigma_{\text{V}})dA_{\text{V}} = \frac{1}{A_{\text{V}}\sqrt{2\pi \sigma^{2}_{\text{V}}}}\exp(\frac{-(ln(A_{\text{V}}/\tilde{A}_{\text{V}}))^{2}}{2\sigma^{2}_{\text{V}}})dA_{\text{V}}
\label{eqn:Av}
\end{equation}
Since $\tilde{m}_{a}$ and $\tilde{m}_{b}$ are not analytic, there is no closed analytic form for the joint probability density function. Therefore, to approximate the likelihood, we note that the likelihood is the expectation of $p(m_{a}|\tilde{m_{a}})$ and $p(m_{b}|\tilde{m}_{b})$ with respect to the initial mass M and extinction $A_{\text{V}}$. As such, we estimate the joint probability density function as:
\setlength{\belowdisplayskip}{5pt}
\begin{align}
    \mathcal{L}(m_{a}, m_{b}| \theta,\sigma_{\text{V}}) 
    &= E_{M, A_{\text{V}}}[p(m_{a}|\tilde{m}_{a}(t, [\text{M}/\text{H}],M),A_{\text{V}})p(m_{b}|\tilde{m}_{b}(t, [\text{M}/\text{H}],M),A_{\text{V}})] \notag \\
    &\approx \frac{1}{N_{M}} \frac{1}{N_{A_{\text{V}}}} \sum_{l}^{N_{M}}\sum_{n}^{N_{A_{\text{V}}}}p(m_{a}|\tilde{m}_{a}(t, [\text{M}/\text{H}],M^{(l)}),A_{\text{V}}^{(n)})p(m_{b}|\tilde{m}_{b}(t, [\text{M}/\text{H}],M^{(l)}),A_{\text{V}}^{(n)})
    \label{eq:PIFestimate}
\end{align}
Where each $M^{(l)}$ is a draw from $p(M)$, and each $A_{\text{V}}^{(n)}$ is a draw from $p(A_{\text{V}} | \tilde{A}_{\text{V}},\sigma_{\text{V}})$. To minimize sampling noise, we used perfect sampling instead of random sampling when drawing from this distribution. Specifically, we used equidistant sequences between 0 and 1 to compute quantiles via the inverse cumulative distribution function, effectively employing a quasi-Monte Carlo approach to approximate the expected value. Figure \ref{fig:Likelitables} presents two examples of the likelihood estimates using HST/ACS F606W vs F814W bands for two different ages.

Real stellar populations will likely consist of a mixture of populations with varying ages and metallicities. Therefore, we propose a mixture data-generating model, which is a weighted sum of the individual joint probability density functions:
\setlength{\belowdisplayskip}{5pt}
\begin{equation}
    \mathcal{L}(m_{A},m_{B}|\{w_{t,z,e}\}_{t=1, z=1, e=1}^{T,Z,E})=\sum_{t}^{T} \sum_{z}^{Z}\sum_{e}^{E}w_{t,z,e}\mathcal{L}(m_{A}, m_{B} | \theta_{t,z,e})
    \label{eq:Mixture Model}
\end{equation}

Where $\{w_{t,z,e}\}$ are the weights for age index t, metallicity index z, and median extinction index $e$. We discretize these parameters for computational efficiency, aligning with the format of the stellar models and isochrones. This study considers a wide range of t, z, e values. Specifically, the ages under consideration span $\log_{10}$(t/yr) = 6.50 to $\log_{10}$(t/yr) = 8.80, where the lower bound reflects the resolution limits of the stellar model, and the upper bound is constrained by the detection limit to ensure quality data. Likewise, [M/H] values range from -0.40 to 0.20, and median extinctions, $\tilde{A}$$_{\text{V}}$, range from 0.0 to 1.50. When extinction is included in the inference, the upper age limit decreases to $\log_{10}$(t/yr) = 8.40, as older stars would fall under the detection limit, as indicated in sections \ref{subsec:SyntheticTest}, \ref{subsec:RealData}. The spacing of these values was chosen with practical considerations in mind, though further refinement is necessary to optimize the grid for inference, which is an effort that goes beyond the scope of this paper. For instance, to assess the impact of discretization, we examined how differences in magnitude coarsely translated to variations in inferred age and mass. Based on our preliminary analysis, we selected the spacing for t, z, and e to balance resolution with computational efficiency.

To establish robust parameter weights, we use a statistical framework, specifically Bayesian inference. Due to the complexity of sampling from the posterior distribution, we implement a Gibbs sampling algorithm \citep{Geman_1984_GibbsSeminole, Gelfand1990_Gibbs, Casella1992_GibbsExplained} with latent variables (labels) $r_{i}$ that assign specific age, metallicity, and median extinction values to each star. Thus, the mixture model in eq.~(\ref{eq:Mixture Model}) can be equivalently expressed as:
\setlength{\belowdisplayskip}{5pt}
\begin{eqnarray}\nonumber
(m_{A,i},m_{B,i} | r_i = (t', z', e'))  &\sim& \mathcal{L}(m_A,m_B|\theta_{t',z', e'}) \\\label{mix_model_ri}
r_i = (t', z', e') | \{w_{t,z,e}\}^{T,Z,E}_{t=1,z=1,e=1} & \sim & w_{t',z',e'} \quad (i = 1...N_{\rm stars}) \, .
\label{eq: Gibbs Mixture Model}
\end{eqnarray}

We conclude model specification by assigning a prior
distribution to the weights. Specifically, we assume that:
\begin{equation}
    \{w_{t,z,e}\}_{t=1,z=1,e=1}^{T,Z,E} \sim {\rm Dirichlet}(1,..,1)
    \label{eq: Dirichlet Weighting}
\end{equation}
Where Dirichlet(1,...,1) describes a $(T \times Z \times E)$-dimensional Dirichlet distribution. Using the model outlined in eq.\ref{eq: Gibbs Mixture Model}, and the prior described above in eq.\ref{eq: Dirichlet Weighting}, the Gibbs sampler iterates between updating the labels $(r_{i})$ and the weights $\{w_{t,z,e}\}$.

To update the labels $(r_{i})$, we sample from the conditional
distribution
\begin{equation}
P(r_i = (t',z',e')| \{ w_{t,z,e} \}^{T,Z,E}_{t=1,z=1,e=1}) = \frac{w_{t',z',e'} p(m_{a,i},m_{b,i} | \theta_{t',z',e'}) }{ \sum_{t} \sum_{z} \sum_e w_{t,z,e} p(m_{a,i},m_{b,i}|\theta_{t,z,e})} \,  \quad (i = 1...N_{\rm stars}) \, .
\end{equation}
To update the weights, $\{w_{t,z,e}\}$, we begin by computing the number of labels $r_{i}$, $i = 1, ...N_{stars}$, that are equal to $(t',z',e')$ denoting the resulting count as $N(t = t', z = z', e = e')$. Finally, we update the weights by sampling from the Dirichlet distribution:
\begin{equation}
    \{ w_{t,z,e} \}^{T,Z,E}_{t=1,z=1,e=1} | \{r_i\}_{i=1}^{N_{\rm stars}} \sim {\rm Dirichlet}[1+N(t=1,z=1,e=1),...,1+N(t=T,z=Z,e=E)] \, .
\end{equation}
These weights represent the end of our inference pipeline, and characterize the age, metallicity, and median extinction of our stellar populations. There are multiple examples of these inferred weights for stellar populations in Section \ref{sec:verification} below.

Several computational choices are relevant when using an MCMC sampling process. For instance, addressing autocorrelation, selecting an appropriate number of MCMC steps, and implementing a burn-in period are crucial for obtaining independent and reliable samples.
For this study, we used 5000 Gibbs MCMC steps to ensure thorough exploration of the parameter space. Empirically, we found that MCMC samples settle into a steady-state distribution around 50 steps, so we choose a burn-in period of 50 steps. To further ensure that the samples are independent, we thinned the chain by retaining every 10th sample. Adjusting parameters like the number of steps, burn-in period, and thinning interval can balance computational cost with sample quality. These parameters are dynamically specified in \textit{Stellar Ages}.
\section{Verification} \label{sec:verification}

This section aims to verify the effectiveness of \textit{Stellar Ages} in inferring stellar ages, metallicities, and extinctions of populations.  We use two approaches.  One is to infer the stellar properties of synthetic data for which we know the truth.  The second is to infer the stellar populations of real data and compare our results to other approaches.

\subsection{Verification using Synthetic Data
\label{subsec:SyntheticTest}}

To evaluate the performance of \textit{Stellar Ages}, we generated synthetic data based on two hypothetical stellar populations with known parameters. This controlled environment allows us to evaluate the accuracy and reliability of our code, ensuring a clear benchmark for comparison.

To generate synthetic data, we randomly sample from the initial mass function, isochrone, and observational uncertainties.  We begin by specifying the age, metallicity, extinction, and number of stars for each stellar population. We then use perfect sampling to draw stars from the Kroupa initial mass function IMF \citep{Kroupa2001, Kroupa2002}, and for most of the masses that we sample, this is also the Salpeter IMF \citep{SalpeterIMF_1955}. Next, we use the isochrones to calculate the blue and red magnitudes, and we adjust for the distance modulus.  Given a model for uncertainty, we sample from a normal distribution, where the width of this normal distribution is calibrated using real data from a particular set of observations.  If an extinction is specified, we randomly draw an extinction ($A_{\text{V}}$) from the log-normal distribution (eq.~\ref{eqn:Av})) and use the extinction scale factor, $f(a)$, to calculate the extinction in each magnitude band.   Finally, we apply observational magnitude limits, retaining only stars within the defined magnitude ranges.

To simulate somewhat realistic conditions, we calibrated the distance modulus and uncertainties in the magnitudes using an initial analysis of data surrounding SN 2004dj. The distance modulus to SN 2004dj is 28 \citep{Williams_2014}.  Furthermore, we adjusted the number of stars in each synthetic population to reflect the expected number density around SN 2004dj at circle radii of 50 pc, 100 pc, and 150 pc. We chose these as representative radii because the stars that formed in a common event remain spatially correlated on physical scales up to $\sim$100 pc during the 100 Myr lifetimes of 4 $\text{M}_{\odot}$ stars, even if the cluster is not gravitationally bound \citep{Bastian_2006}. However, as in \citet{Gogarten_2009}, one may reduce this radius to balance including as many coeval stars as possible while limiting contamination. Therefore, we present a wide range of circle radii to reflect various stellar contexts. This tailored approach closely models the observational conditions encountered in actual stellar environments, providing a more realistic benchmark for inferring stellar ages, metallicities, and extinctions as a function of circle radii.

\begin{figure}[H]
    \centering
    \begin{minipage}[b]{0.32\textwidth}
        \centering
        \includegraphics[width=\textwidth]{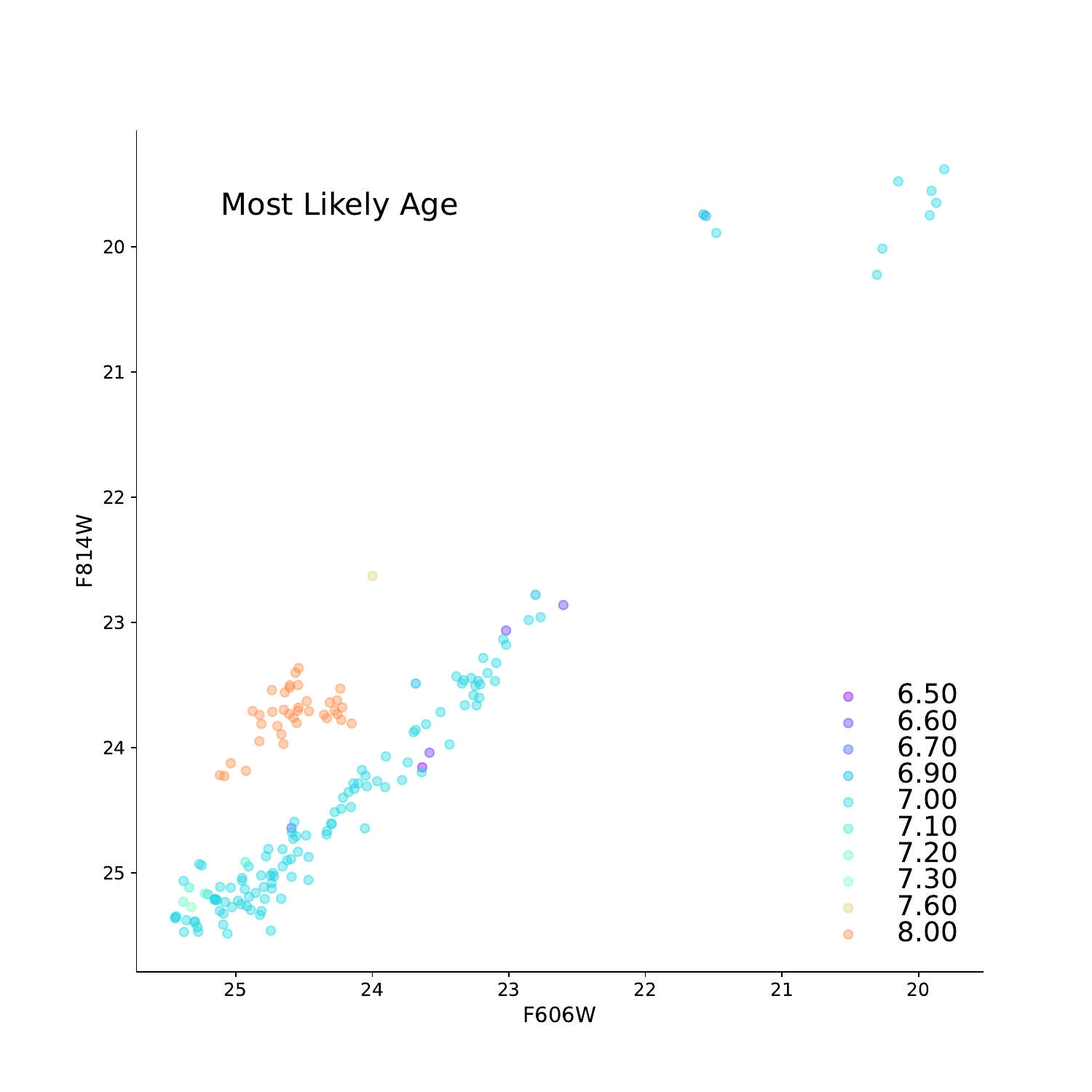}
    \end{minipage}
    \hfill
    \begin{minipage}[b]{0.32\textwidth}
        \centering
        \includegraphics[width=\textwidth]{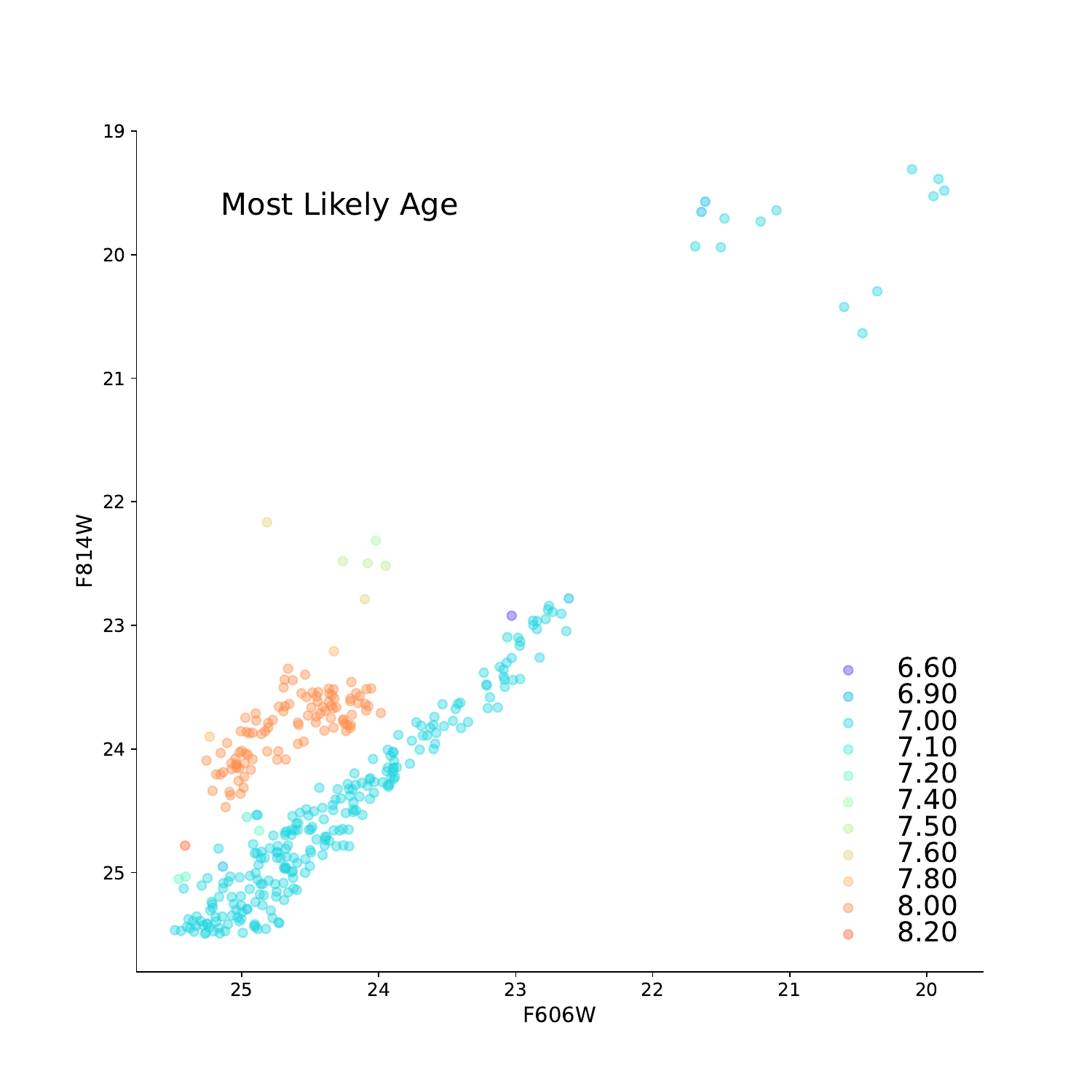}
    \end{minipage}
    \hfill
    \begin{minipage}[b]{0.32\textwidth}
        \centering
        \includegraphics[width=\textwidth]{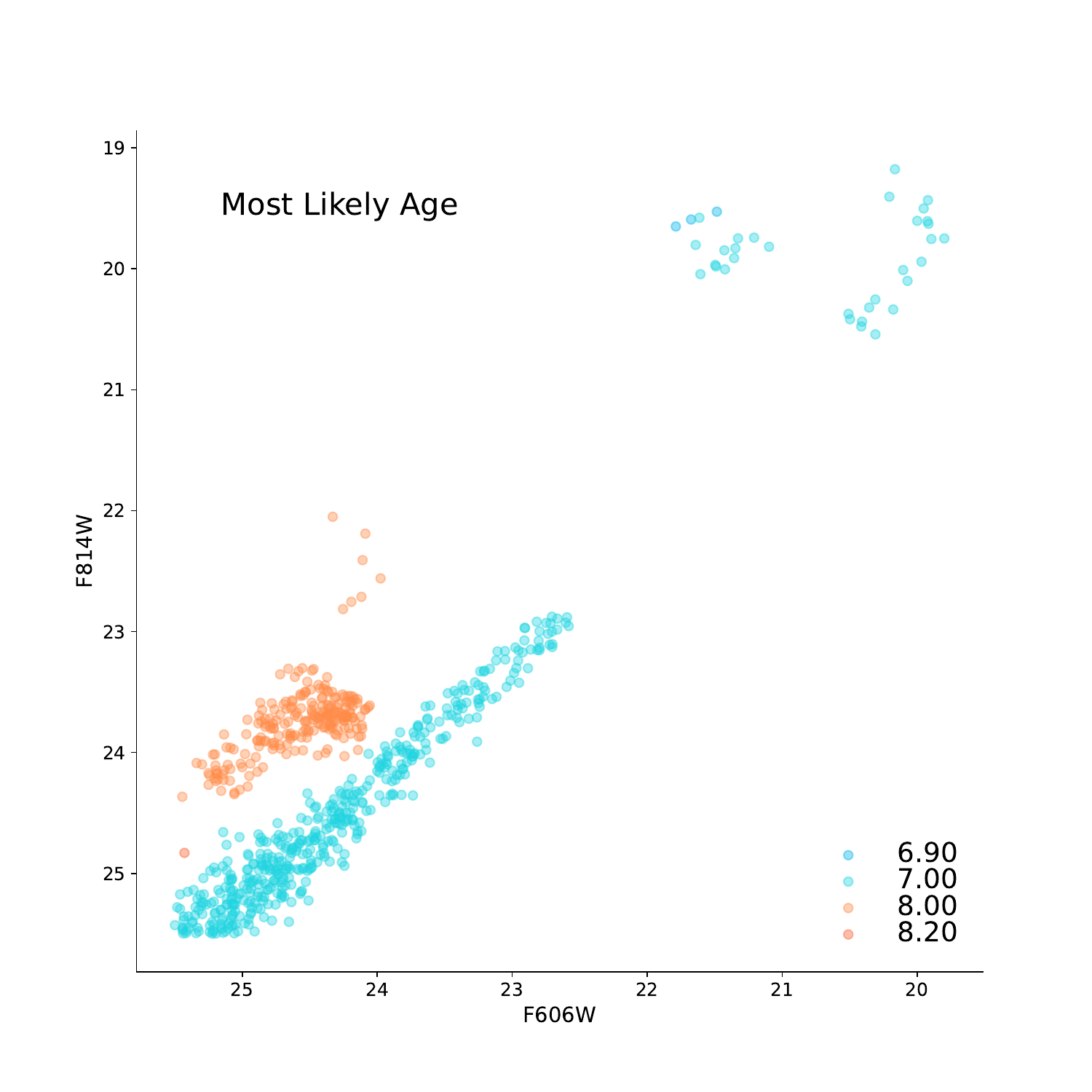}
    \end{minipage}
    \caption{\small F606W vs F814W Magnitude-Magnitude plots for synthetic data. The dots are color-coded by the inferred most likely age of each star. The legend shows the corresponding inferred age for each color. From left to right, the plots represent simulated 50pc, 100pc, and 150pc number densities. The true values of these synthetic populations are $\log_{10}$(t/yr)= $7.0$ and $8.0$, $[M/H] = 0.0$ and $\tilde{A}_{\text{V}} = 0.0$}
    \label{fig:SynthCMDMostLikely}
\end{figure}

After generating synthetic data, we run the \textit{Stellar Ages} inference pipeline, as described in Section \ref{sec:technique}, and we compare the results with the true parameters. Users can select several inference modes within \textit{Stellar Ages}, such as the ``tza'' mode, which infers age, metallicity and extinction, or the ``tz'' mode, which infers only age and metallicity. For completeness, we present synthetic data including extinction in ``tza'' mode. For real data analysis, see Section \ref{subsec:RealData}, where we use both modes to illustrate the impact of marginalizing over extinction, as shown in Figures \ref{fig:10BrightestTZ} and \ref{fig:10BrightestTZA}. Additional inference modes are being developed, such as the ``tzw'' mode, which includes stellar rotation \citep{murphy2024}. As explained in Section \ref{sec:technique}, we initialize the weights using a Dirichlet distribution and iteratively update them with a Gibbs MCMC sampler based upon the likelihood of the observed data given the model parameters. We calculate likelihoods using pre-computed tables, as shown in Figure \ref{fig:Likelitables}, which relate observed magnitudes to the model parameters. These tables are loaded and used to compute the probability of the observed data for each combination of parameters. The combination of synthetic data generation and weight inference is shown in Figure \ref{fig:SynthCMDMostLikely}. We fit the synthetic data for F606W vs F814W in the range 16.5 $<$ F814W and F606W $<$ 25.5.

Figure \ref{fig:SynthCMDMostLikely} shows the magnitude-magnitude diagrams for two synthetic populations at the expected number densities for 50pc, 100pc, and 150pc circle radii. Visual inspection reveals that \textit{Stellar Ages} effectively distinguishes between the populations. The younger population, marked by blue points, clearly displays the MS, Main Sequence Turn-off (MSTO), and Red and Blue Supergiants (BSGs). In contrast, the older population, shown in orange, lacks the MS and only shows Red Supergiants (RSGs). Despite the absence of MS information for the older group, \textit{Stellar Ages} accurately identifies its age. As the circle radii increase from 50pc to 150pc, the inferred ages become more consistent, indicating that the higher number density reduced the degeneracy in age inference.

\begin{figure}[H]
    \centering
    \begin{minipage}[b]{0.32\textwidth}
        \centering
        \includegraphics[width=\textwidth]{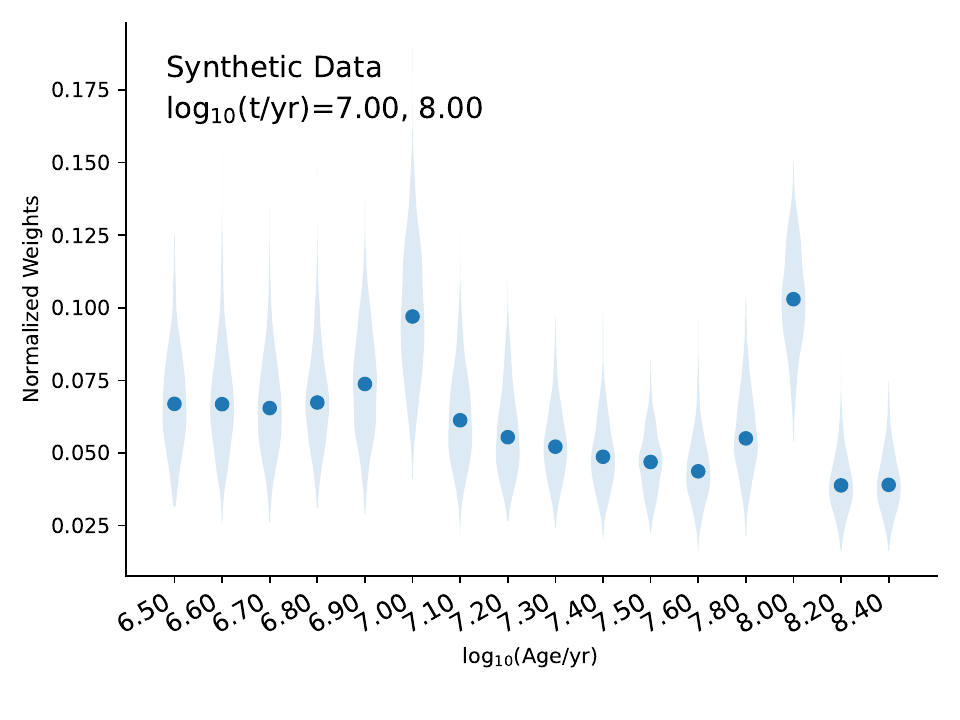}
    \end{minipage}
    \hfill
    \begin{minipage}[b]{0.32\textwidth}
        \centering
        \includegraphics[width=\textwidth]{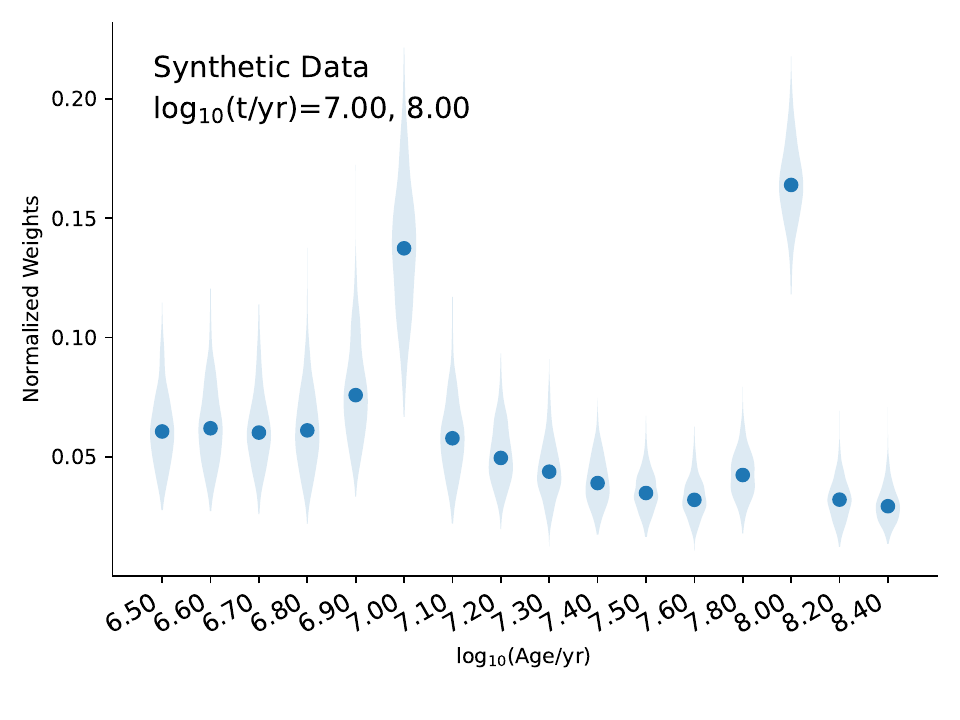}
    \end{minipage}
    \hfill
    \begin{minipage}[b]{0.32\textwidth}
        \centering
        \includegraphics[width=\textwidth]{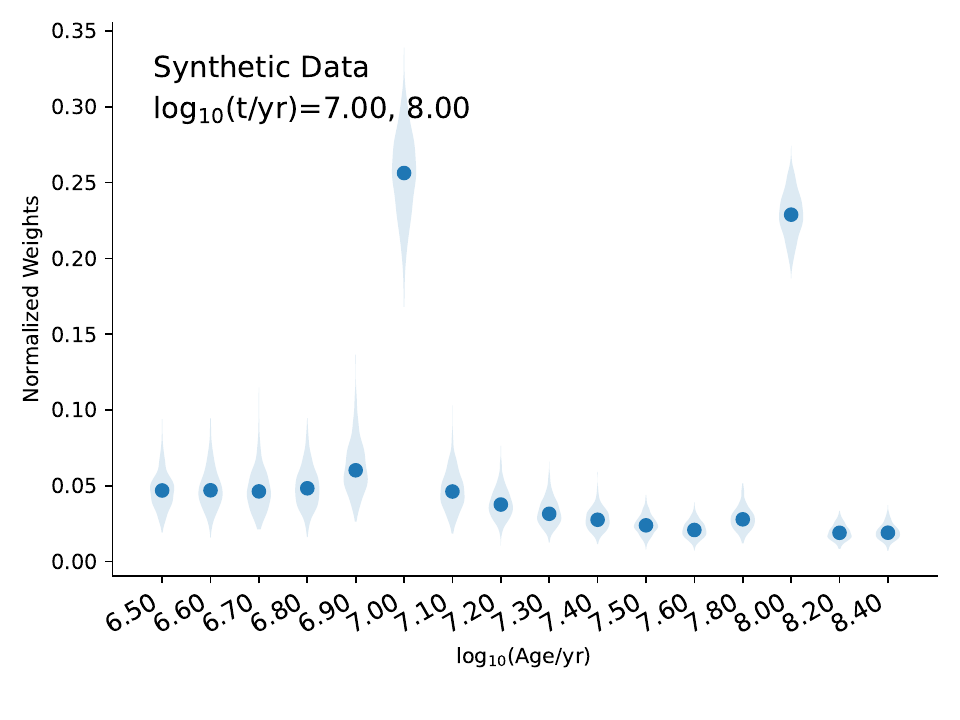}
    \end{minipage}
    \caption{\small Inferred age weights for the synthetic data found from Figure \ref{fig:SynthCMDMostLikely}, marginalized over metallicity and extinction. The panels correspond to the simulated 50pc, 100pc, 150pc datasets. The true values are $\log_{10}$(t/yr) = 7.00, 8.00, $\tilde{A}_{\text{V}} = 0.0, 0.0$ and $[M/H] = 0.00, 0.00$. }
    \label{fig:SynthAgeViolin}
\end{figure}

\begin{figure}[H]
    \centering
    \begin{minipage}[b]{0.32\textwidth}
        \centering
        \includegraphics[width=\textwidth]{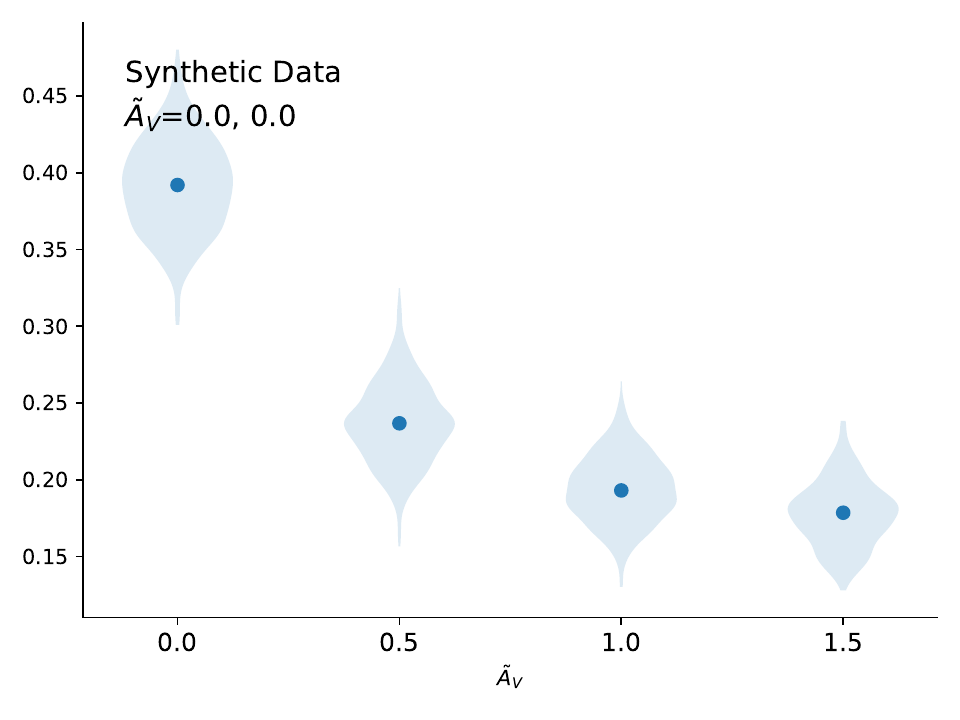}
    \end{minipage}
    \hfill
    \begin{minipage}[b]{0.32\textwidth}
        \centering
        \includegraphics[width=\textwidth]{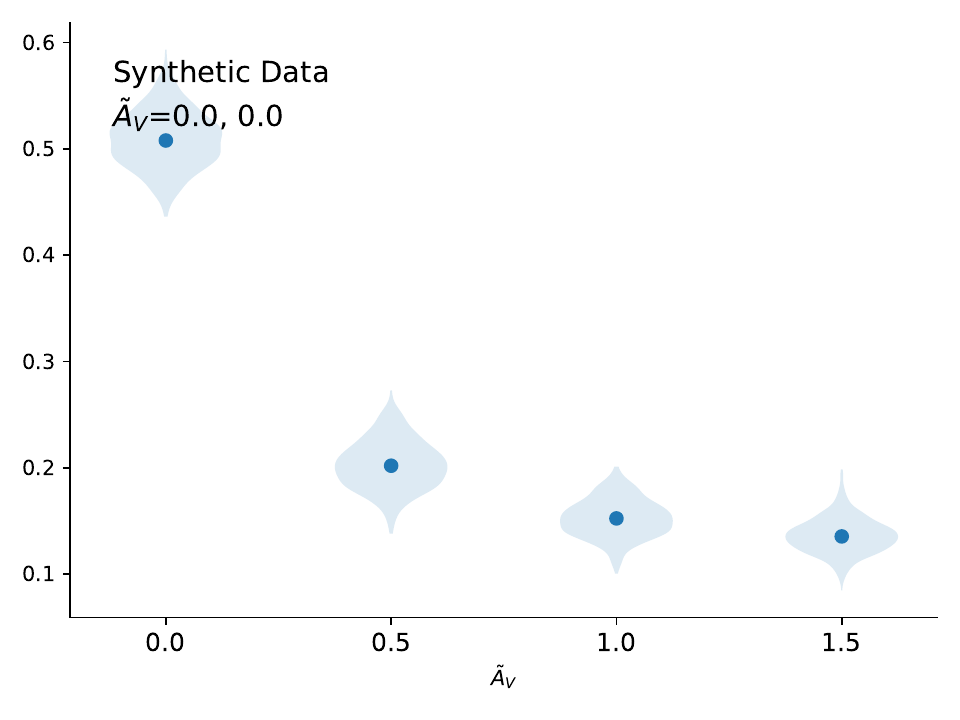}
    \end{minipage}
    \hfill
    \begin{minipage}[b]{0.32\textwidth}
        \centering
        \includegraphics[width=\textwidth]{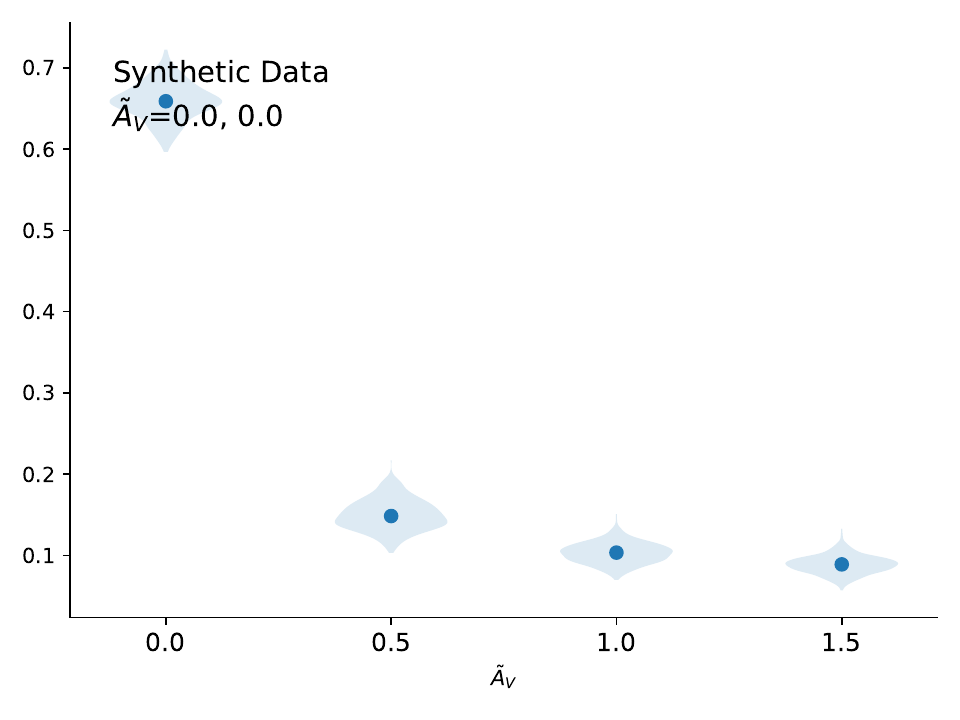}
    \end{minipage}
    \caption{\small Inferred median extinction distribution weights for the synthetic data found in Figure\ref{fig:SynthCMDMostLikely}, marginalized over age and metallicity. The panels present the same simulated 50pc, 100pc, and 150pc datasets as in \ref{fig:SynthAgeViolin}. The true extinctions values were $\tilde{A}_{\text{V}} = 0.0$.}
    \label{fig:SynthAvViolin}
\end{figure}

Figures \ref{fig:SynthAgeViolin} and \ref{fig:SynthAvViolin} further analyze the synthetic data by marginalizing over the age and extinction weights, presenting the most likely ages and extinctions for these groups.
Figure \ref{fig:SynthAgeViolin} shows a bimodal distribution for the age weights across all datasets. While the most likely ages correspond to the true synthetic values, there remains some uncertainty in the 50 pc and 100 pc datasets. The 150 pc dataset recovers both synthetic ages, reflecting the observed spread of inferred ages for Figure \ref{fig:SynthCMDMostLikely} and highlighting the importance of number density (or circle radii) when interpreting age uncertainties.

Figure \ref{fig:SynthAvViolin} presents the inferred median extinction weights for each dataset. Despite a relatively broad extinction distribution, \textit{Stellar Ages} successfully recovers the true extinction value. Here, the extinction degeneracy is lifted by the 100 pc dataset, perhaps reflecting the coarser grid size in the extinction weights.

Our results demonstrate that \textit{Stellar Ages} accurately recovers the known values of age, metallicity, and extinction from the synthetic datasets, providing an initial proof of concept. We next apply the code to real observational data surrounding SN 2004dj to test its performance in practical scenarios where the true characteristics of the stellar population are informed by prior analyses, but not directly constrained.

\subsection{Verification using Real Data} 
\label{subsec:RealData}

We selected SN 2004dj for our analysis due to its proximity and well-constrained extinction and mass uncertainties \citep{Williams_2014, Wang_2005}. Additionally, its host galaxy, NGC 2403, was imaged before the supernova as part of the Beijing-Arizona-Taiwan-Connecticut (BATC) Multicolor Sky Survey from 1995 to 2003, providing excellent precursor photometry, for comparison. This makes SN 2004dj an ideal candidate for testing our methodology.

\begin{figure}[H]
    \centering
    \begin{minipage}[b]{0.32\textwidth}
        \centering
        \includegraphics[width=\textwidth]{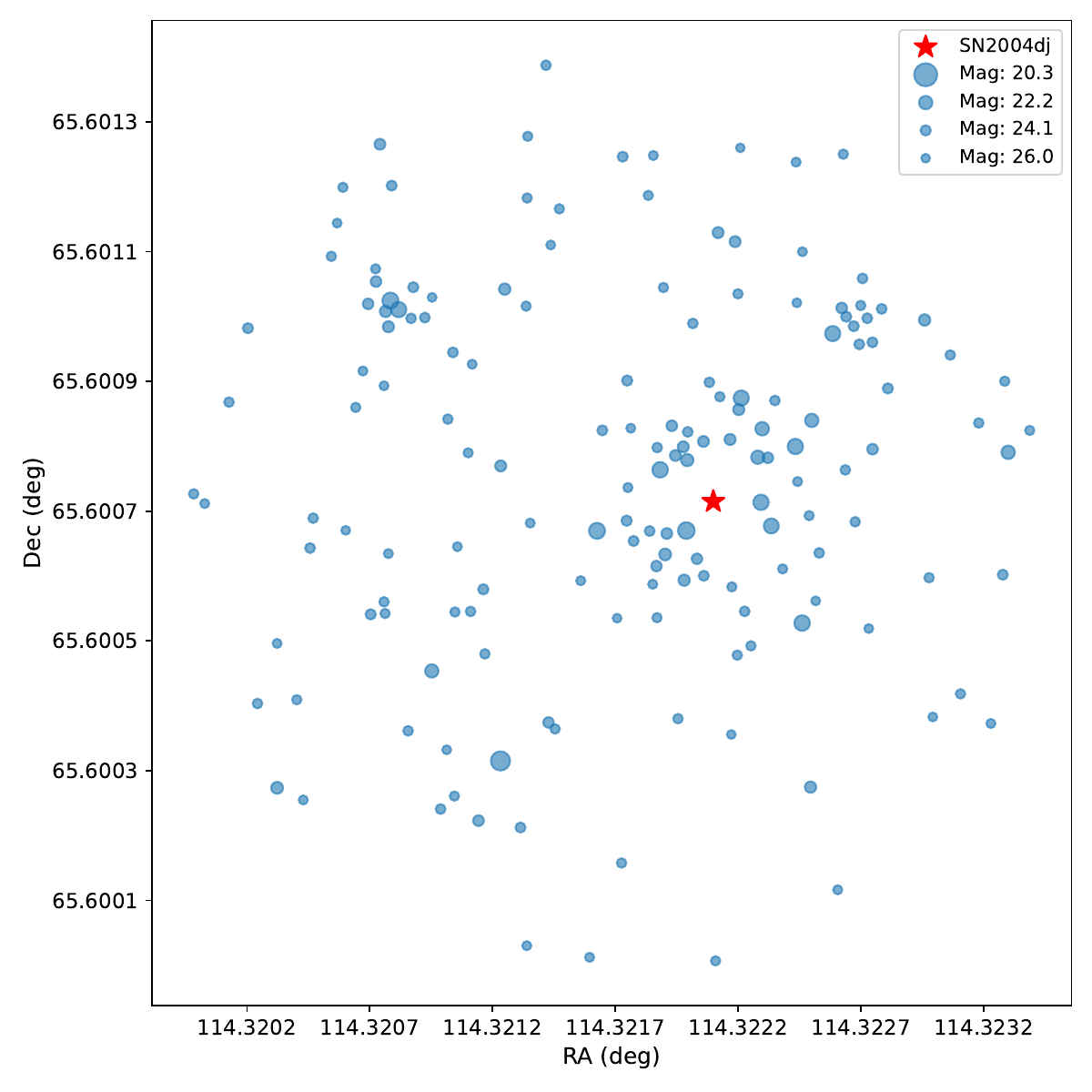}
    \end{minipage}
    \hfill
    \begin{minipage}[b]{0.32\textwidth}
        \centering
        \includegraphics[width=\textwidth]{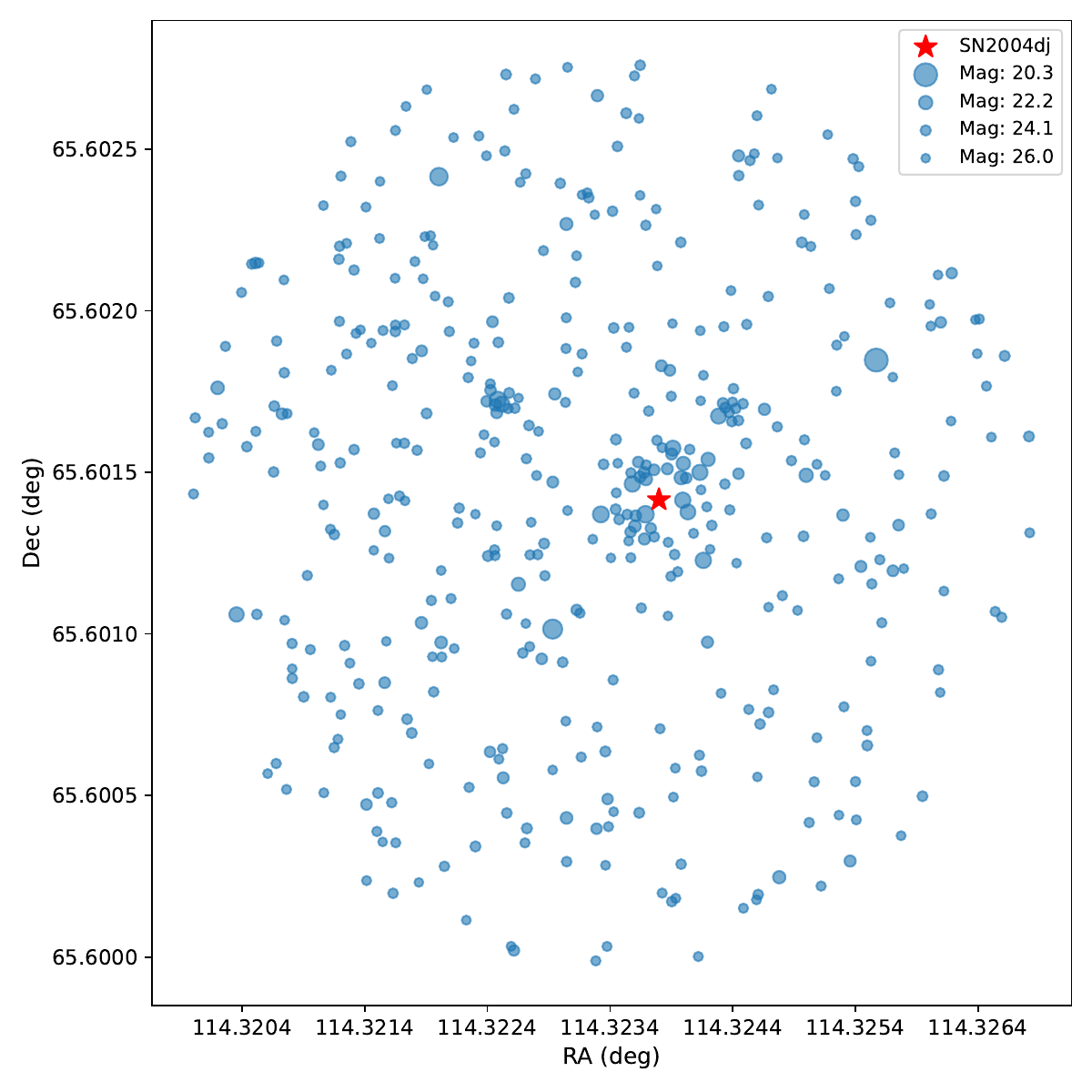}
    \end{minipage}
    \hfill
    \begin{minipage}[b]{0.32\textwidth}
        \centering
        \includegraphics[width=\textwidth]{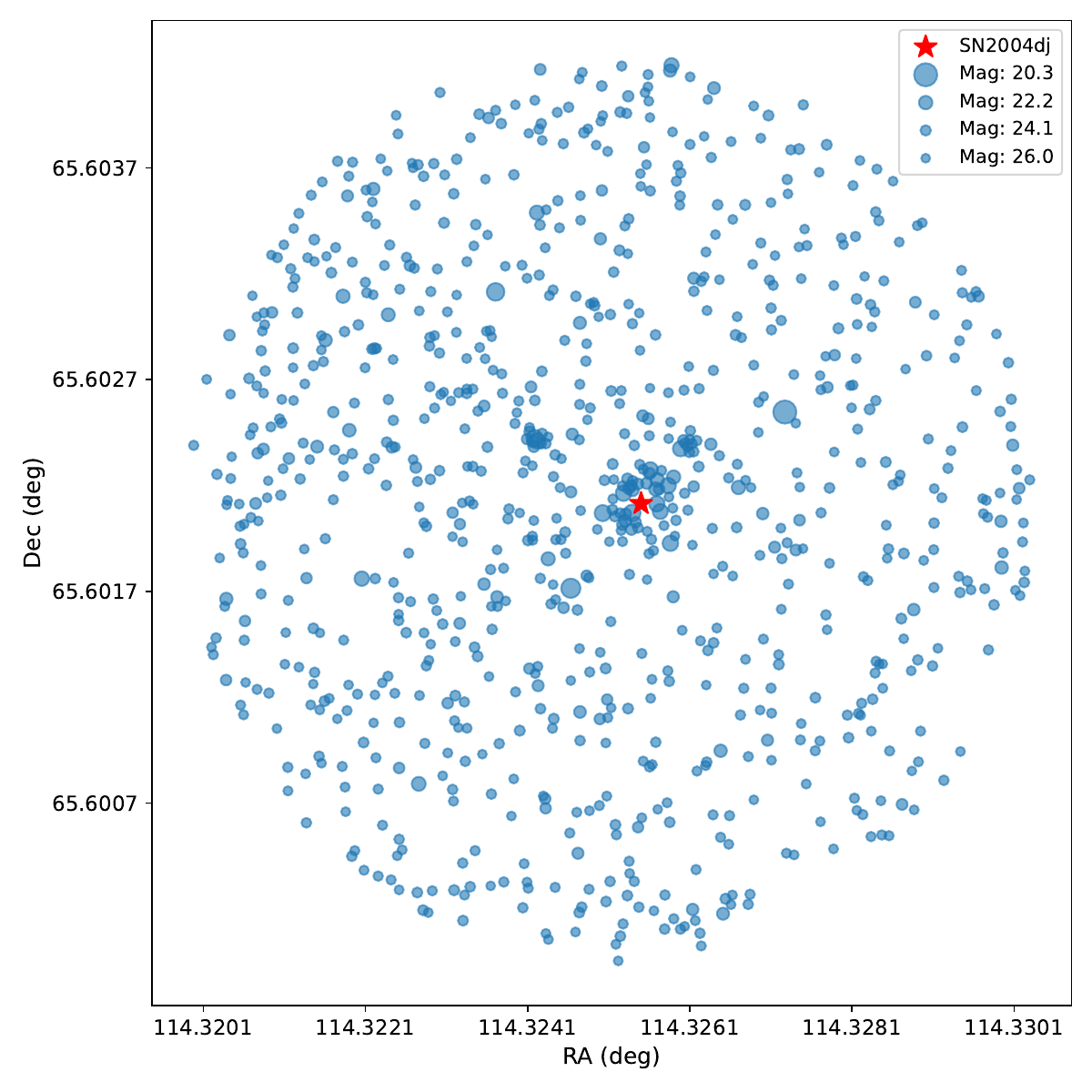}
    \end{minipage}
    \caption{\small RA and DEC coordinates for all stars near SN2004dj that have sufficient quality. As in figures \ref{fig:SynthCMDMostLikely} through \ref{fig:SynthAvViolin}, the panels are organized by the circle radii surrounding SN2004dj. The left panel presents all stars within 50 pc, the middle panel shows all stars within 100 pc, and the right panel shows all within 150 pc. The sizes of the points reflect the relative magnitude of the star across all panels. The red star indicates the location of SN2004dj.}
    \label{fig:SkyCoord}
\end{figure}

We analyzed stellar data surrounding SN 2004dj within three circles centered on SN 2004dj. The radii of these circles correspond to projected distances of 50 pc, 100 pc, and 150 pc.  The data within these circles allow us to test the robustness of the conclusions.  We retrieved archival HST data from proposal ID 10607 (PI: Sugerman). All the data used in this paper can be found in MAST\dataset[https://doi.org/10.17909/hae1-qj98]{https://doi.org/10.17909/hae1-qj98}. This dataset met two key criteria: sufficient exposure time ($\geq$ 1000 s) and coverage in two distinct wavelength bands.
After acquiring the data, we performed photometry using DOLPHOT \citet{Dolphot2016}, a modified version of HSTphot \citep{Dolphin_2000} optimized for ACS. We applied strict criteria for stars, choosing those with signal-to-noise ratios (S/Ns) $>$ 4, sharpness squared $<$ 0.15, and crowding $<$ 1.3 as in \citet{diaz-rodriguez2018, Murphy_2018}. Using this high-quality data, we created Figure \ref{fig:SkyCoord}, which shows the right ascension versus declination of the stars surrounding SN 2004dj.  The sizes of the points is proportional to the magnitude of the stars.

\begin{figure}[H]
    \centering
    \begin{minipage}[b]{0.32\textwidth}
        \centering
        \includegraphics[width=\textwidth]{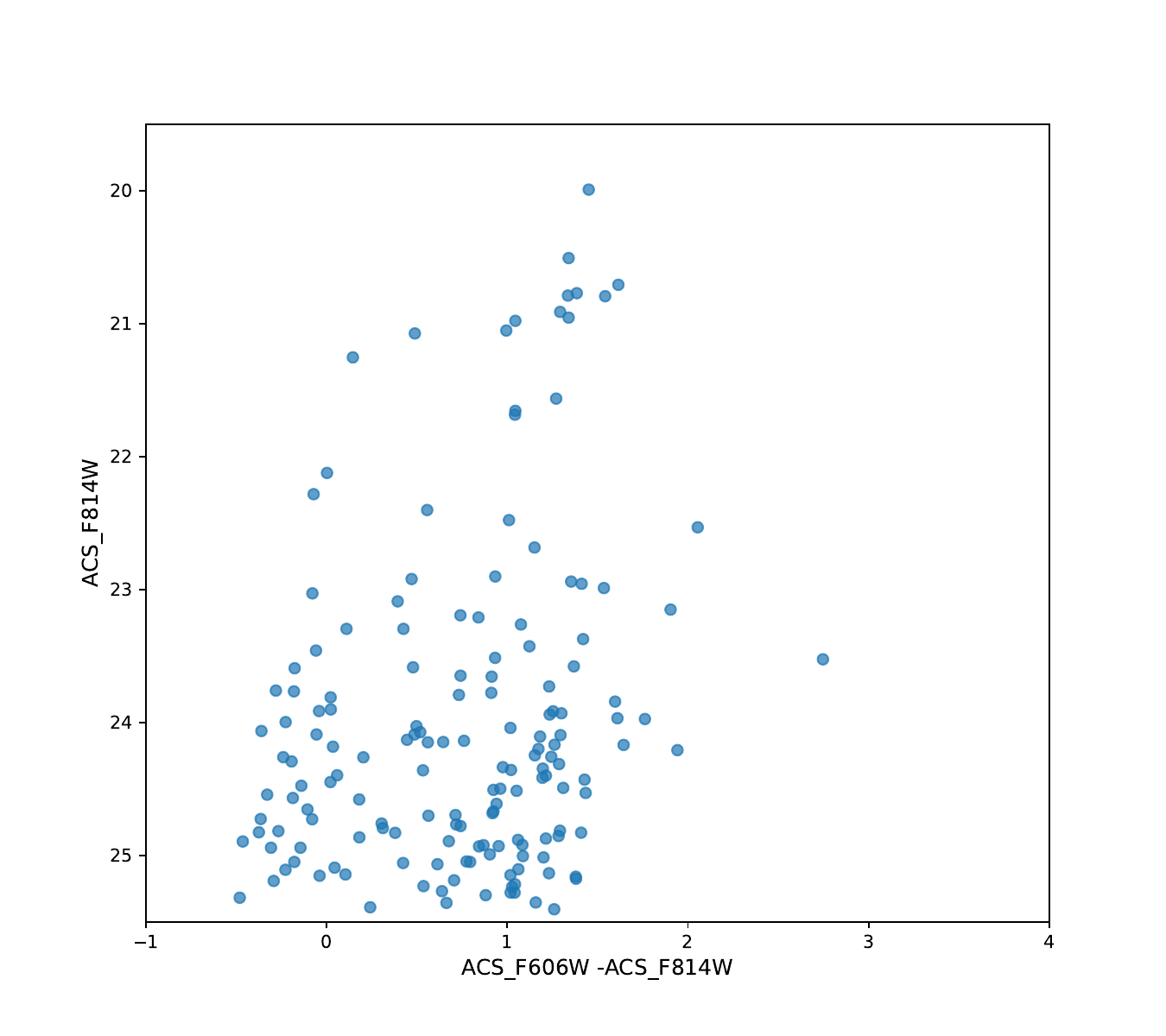}
    \end{minipage}
    \hfill
    \begin{minipage}[b]{0.32\textwidth}
        \centering
        \includegraphics[width=\textwidth]{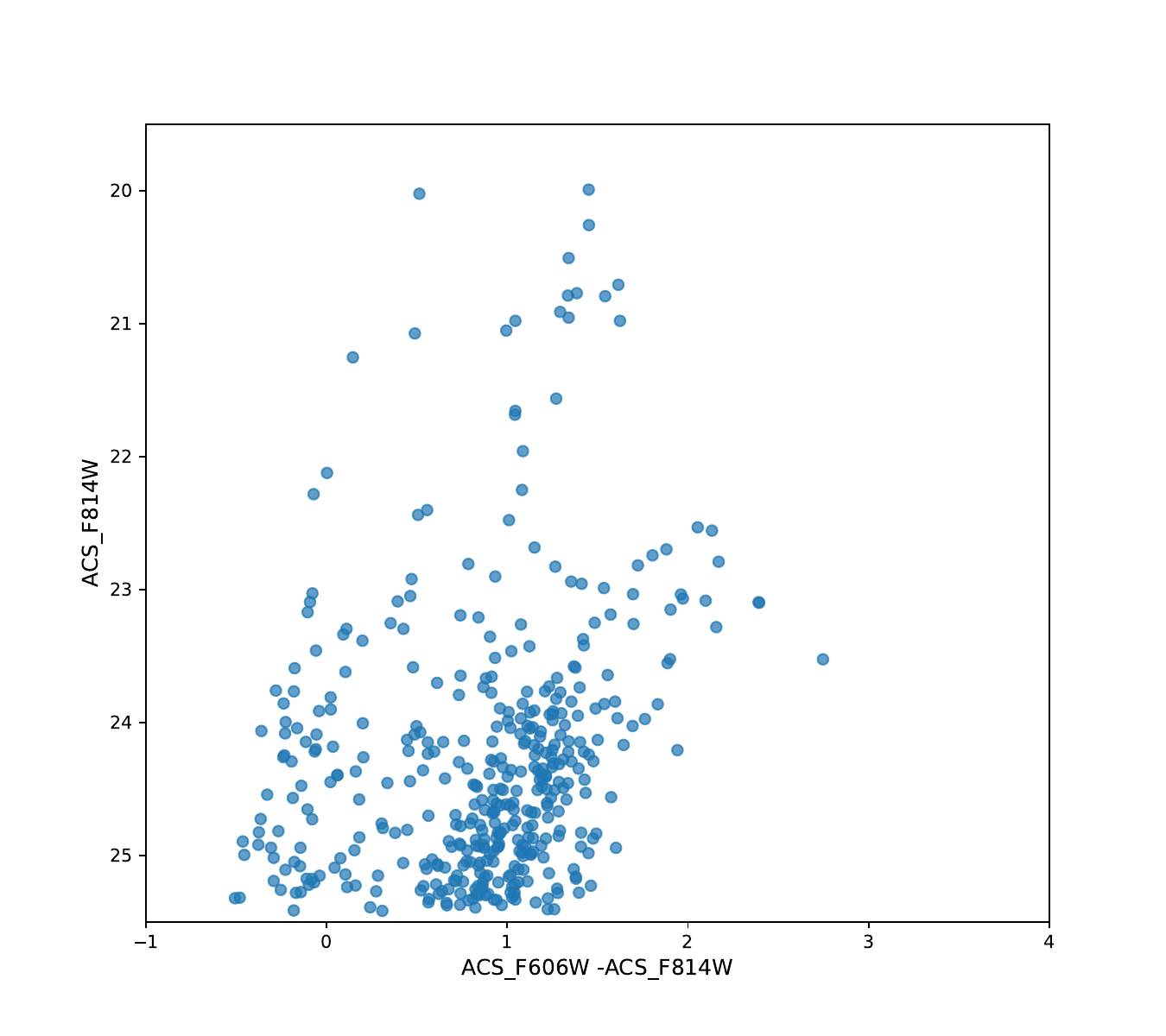}
    \end{minipage}
    \hfill
    \begin{minipage}[b]{0.32\textwidth}
        \centering
        \includegraphics[width=\textwidth]{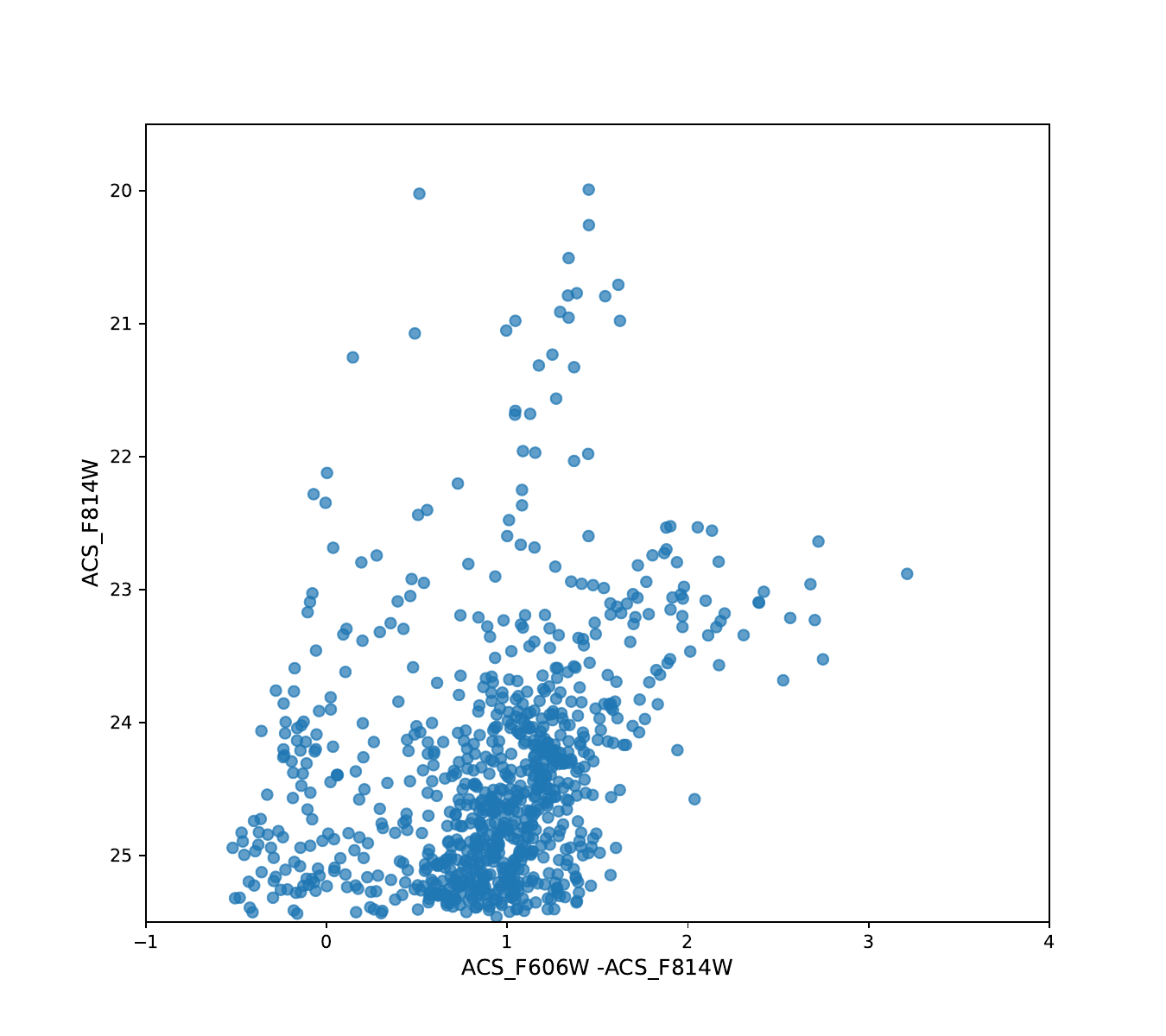}
    \end{minipage}
    \par\vspace{-1.02em}
    \begin{minipage}[b]{0.32\textwidth}
        \centering
        \includegraphics[width=\textwidth]{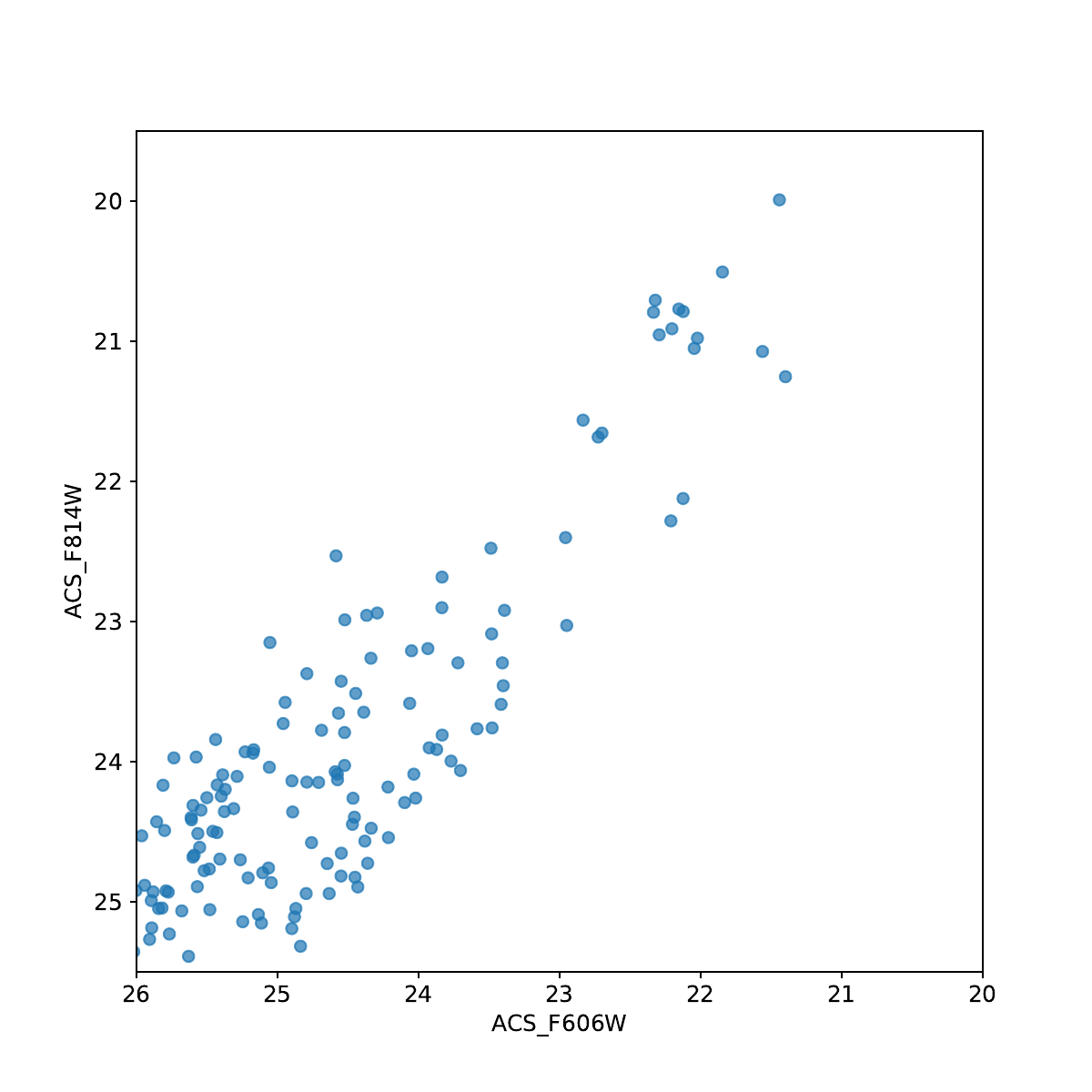}
    \end{minipage}
    \hfill
    \begin{minipage}[b]{0.32\textwidth}
        \centering
        \includegraphics[width=\textwidth]{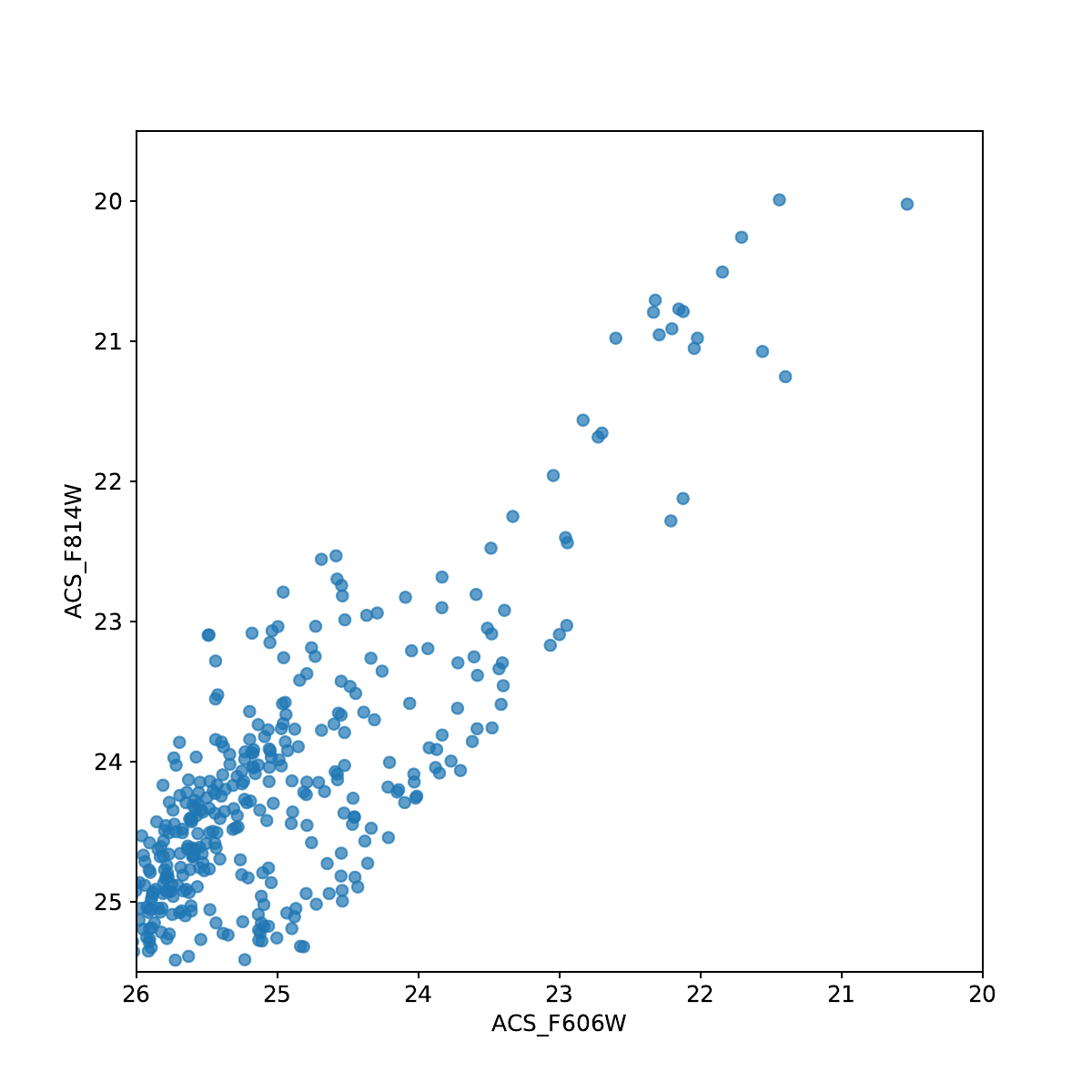}
    \end{minipage}
    \hfill
    \begin{minipage}[b]{0.32\textwidth}
        \centering
        \includegraphics[width=\textwidth]{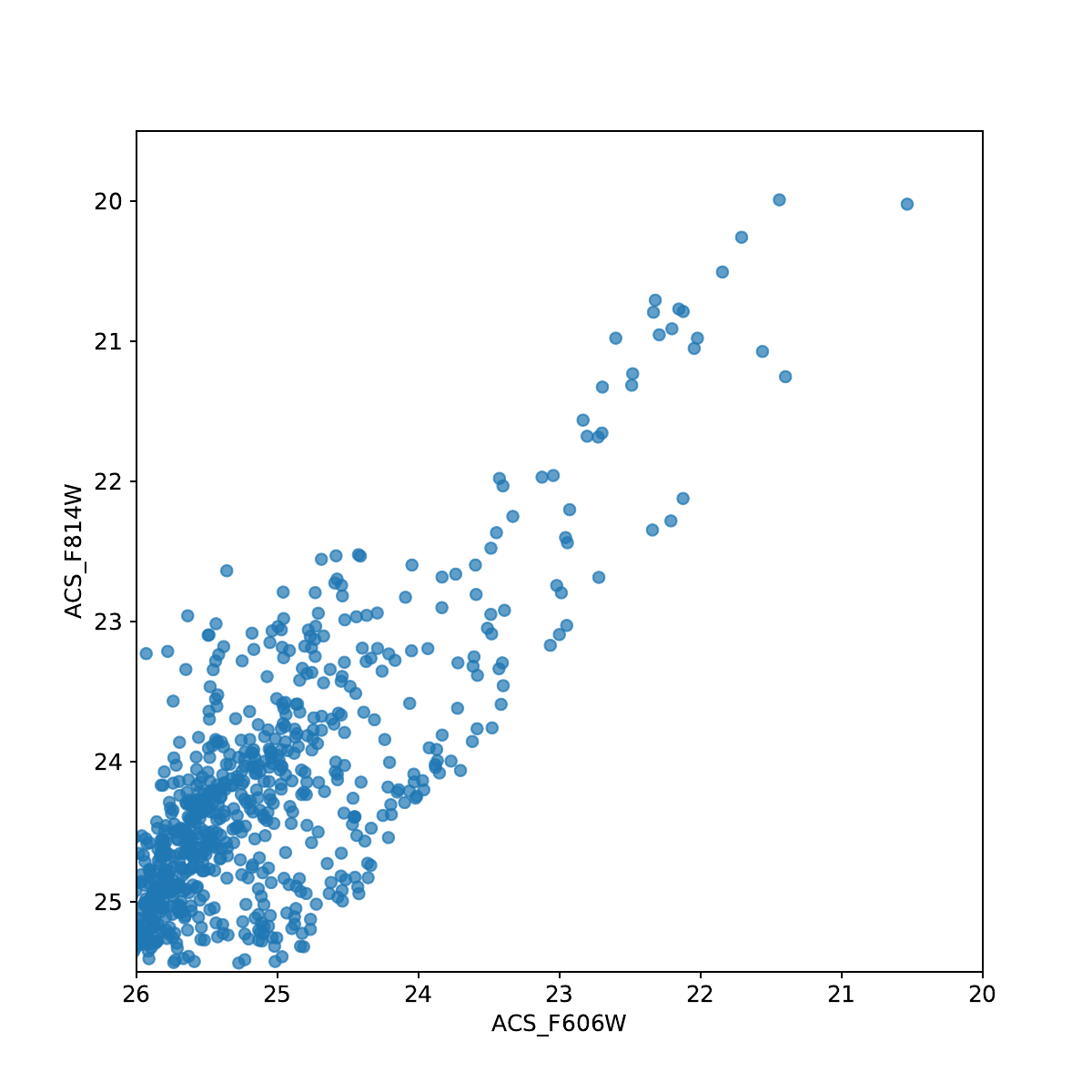}
    \end{minipage}
    \caption{\small Top row: Color magnitude diagrams (CMD) of all stars with sufficient quality near SN2004dj, Bottom row: magnitude-magnitude diagrams of all stars with sufficient quality near SN2004dj. These plots are organized by circle radii. From the left column to the right, represents the 50pc, 100pc and 150pc data surrounding the SN location.}
    \label{fig:CMD_MagMag}
\end{figure}

Next, Figure~\ref{fig:CMD_MagMag} presents the CMDs and magnitude-magnitude diagrams for the same photometry in Figure \ref{fig:SkyCoord}. The top panel of Figure~\ref{fig:CMD_MagMag} plots F606W-F814W versus F814W, and the bottom panel shows F606W versus F814W.

\begin{figure}[H]
    \centering
    \begin{minipage}[b]{0.32\textwidth}
        \centering
        \includegraphics[width=\textwidth]{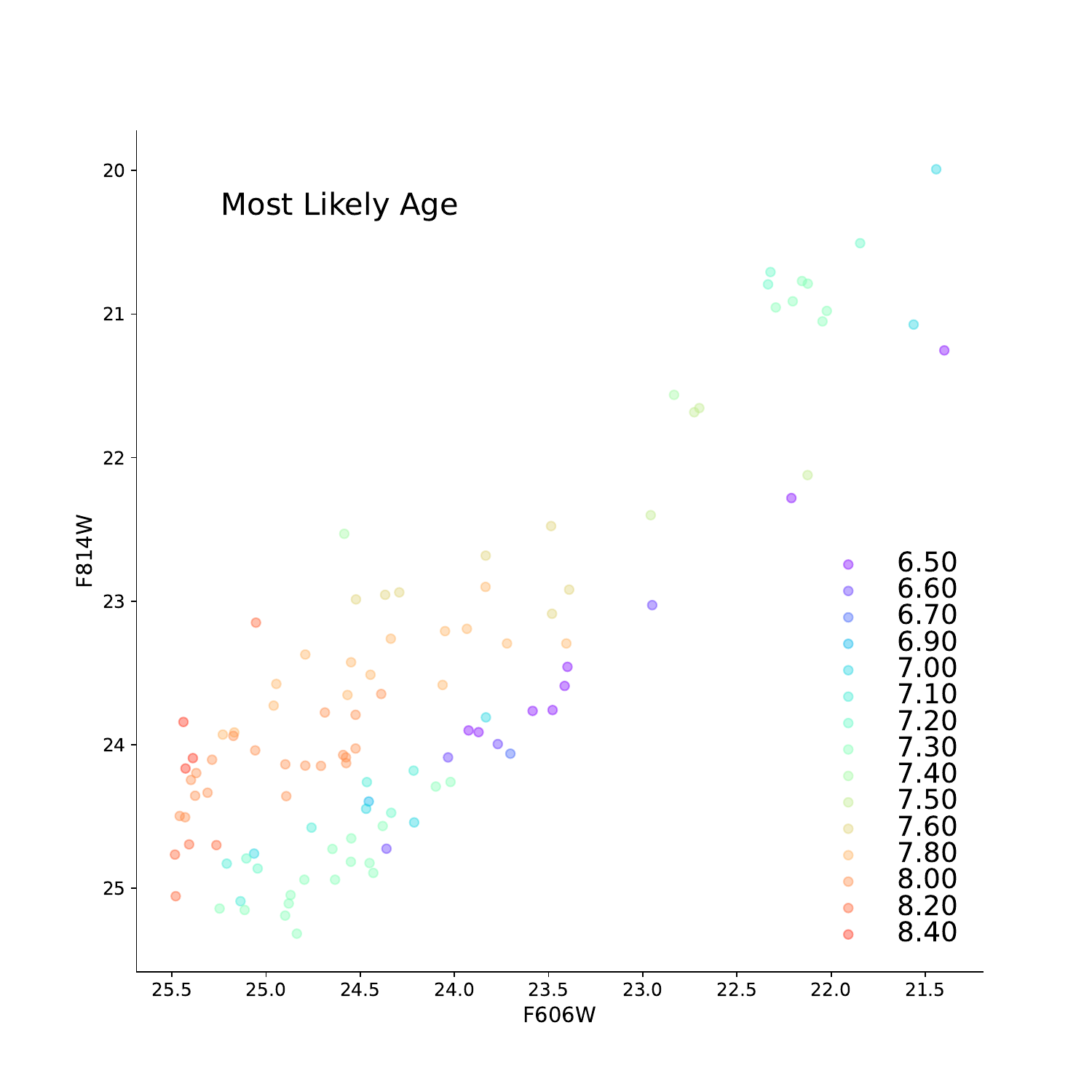}
    \end{minipage}
    \hfill
    \begin{minipage}[b]{0.32\textwidth}
        \centering
        \includegraphics[width=\textwidth]{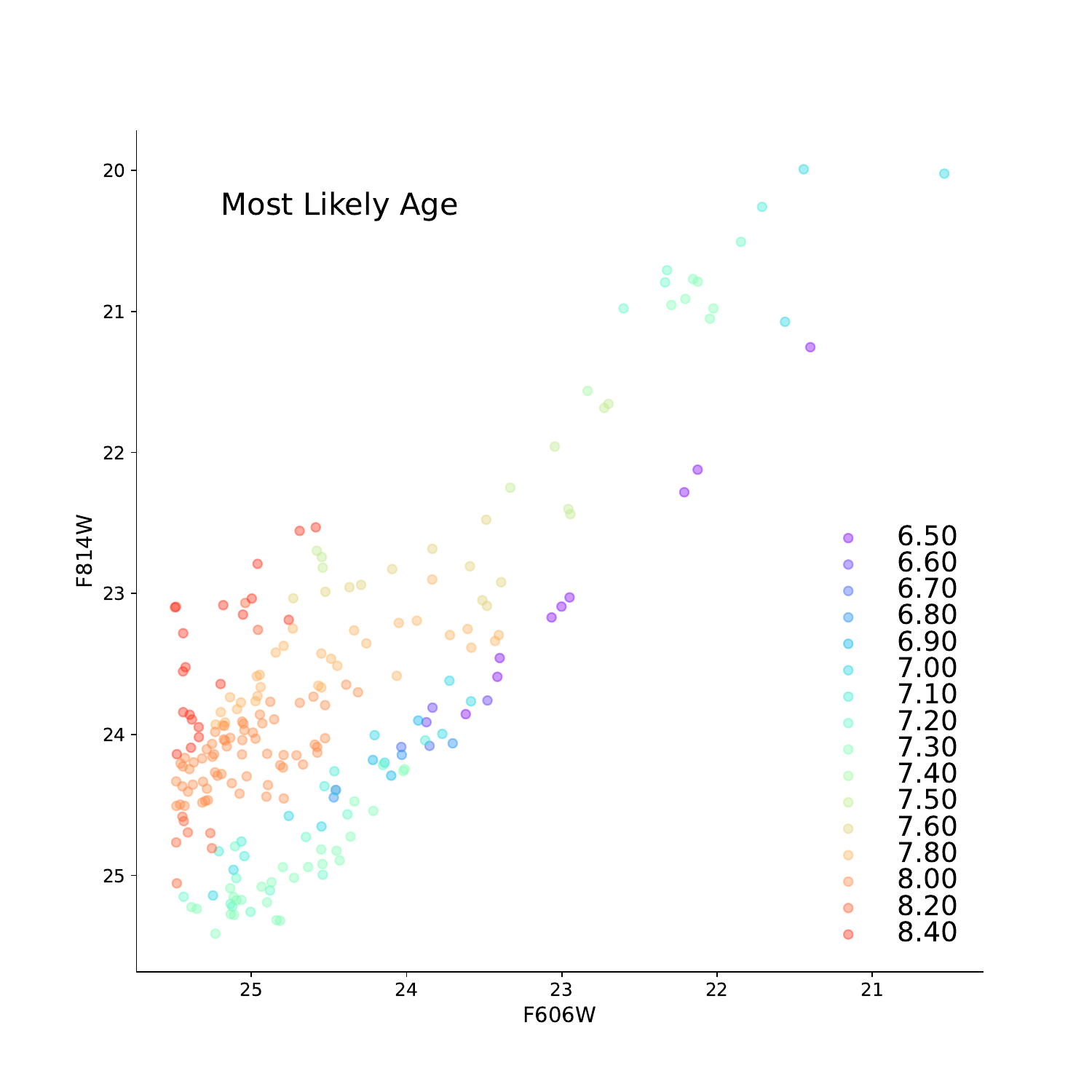}
    \end{minipage}
    \hfill
    \begin{minipage}[b]{0.32\textwidth}
        \centering
        \includegraphics[width=\textwidth]{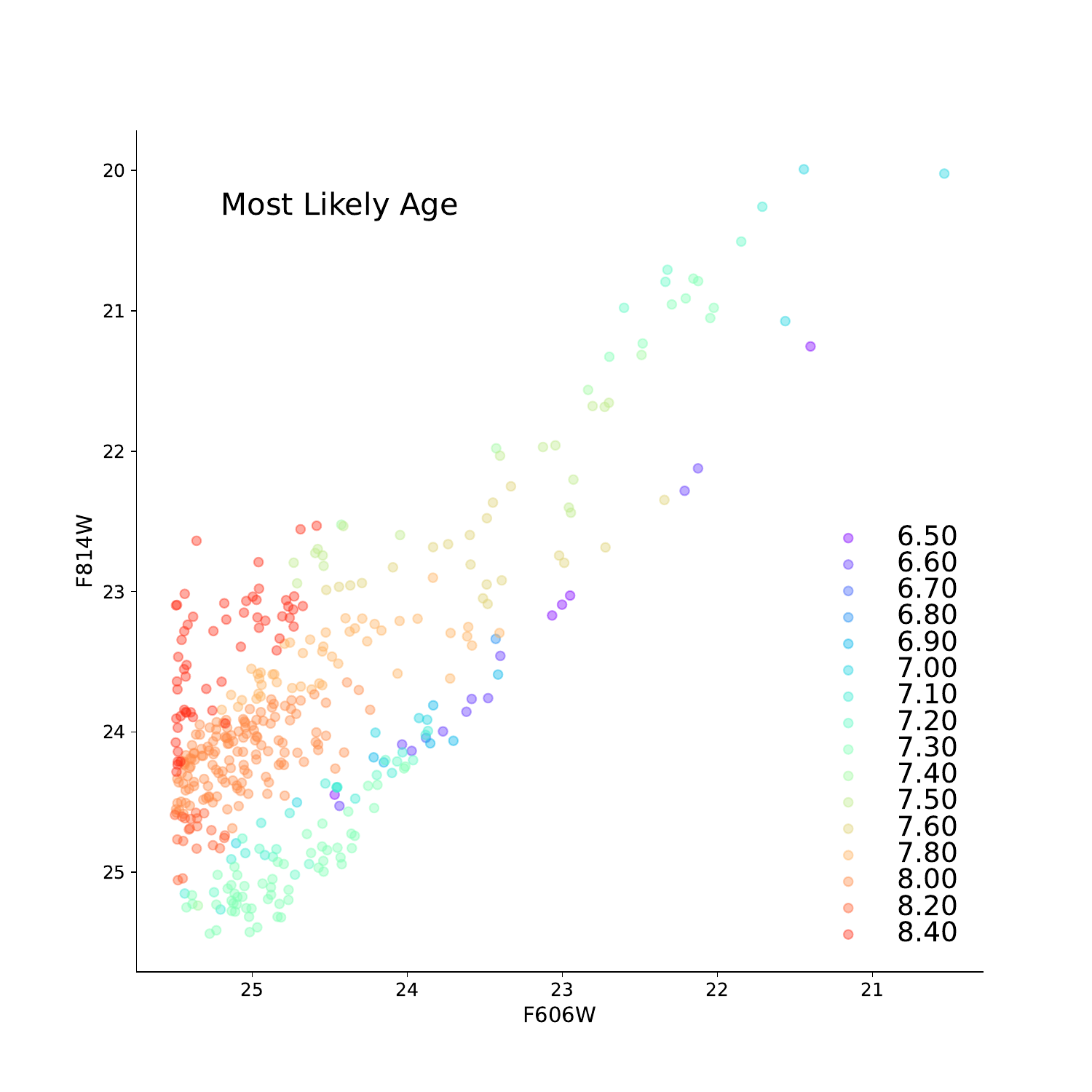}
    \end{minipage}
    \caption{\small Mag-Mag plots and the most likely inferred age for each star. The color of the dots represents the most likely inferred age for each star, marginalizing over metallicity and extinction. Similar to the synthetic test in figure \ref{fig:SynthCMDMostLikely}, the allowed $\tilde{A}_{\text{V}}$ values are 0.0, 0.5, 1.0, and 1.5, and allowed $[M/H]$ values are -0.40, -0.20, 0.00, and 0.20. The panels are organized by circle radii considered from SN2004dj. I.e., the left panel shows the 50pc dataset, the middle panel shows the 100pc dataset, and the right panel shows the 150pc dataset.}
    \label{fig:CMDMostLikely}
\end{figure}

Figure \ref{fig:CMDMostLikely} shows magnitude-magnitude (F606W versus F814W) diagrams, where the points are color-coded to represent the inferred most likely ages of each star. By marginalizing over metallicity and extinction, we identify trends in the age distribution of the stellar population. For example, some blue stragglers appear along the MS. Notably, the blue points ($\log_{10}$(t/yr) $\sim$ 7.30) form a distinct MS with corresponding RSGs. In contrast, a younger population represented by purple points ($\log_{10}$(t/yr) $\sim$ 6.60) lacks a clearly defined MS. This absence of a MS and corresponding evolved stars for this age could suggest incompleteness in the stellar models. Additional factors, such as rotation and binarity, may need to be incorporated to improve the accuracy of age inferences for these stars.

\begin{figure}[H]
    \centering
    \begin{minipage}[b]{0.32\textwidth}
        \centering
        \includegraphics[width=\textwidth]{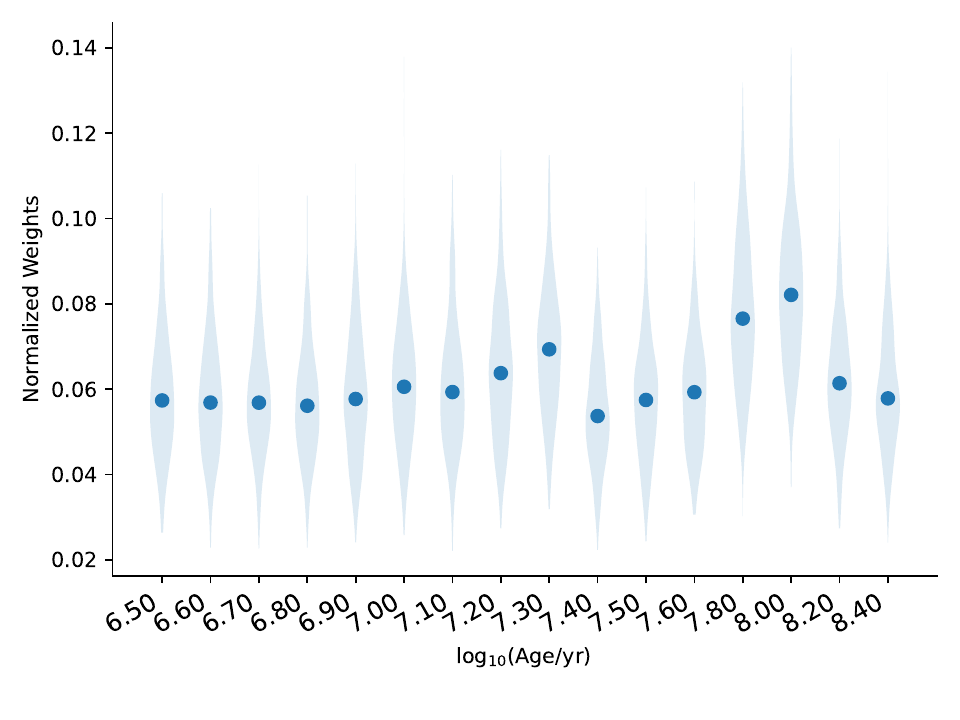}
    \end{minipage}
    \hfill
    \begin{minipage}[b]{0.32\textwidth}
        \centering
        \includegraphics[width=\textwidth]{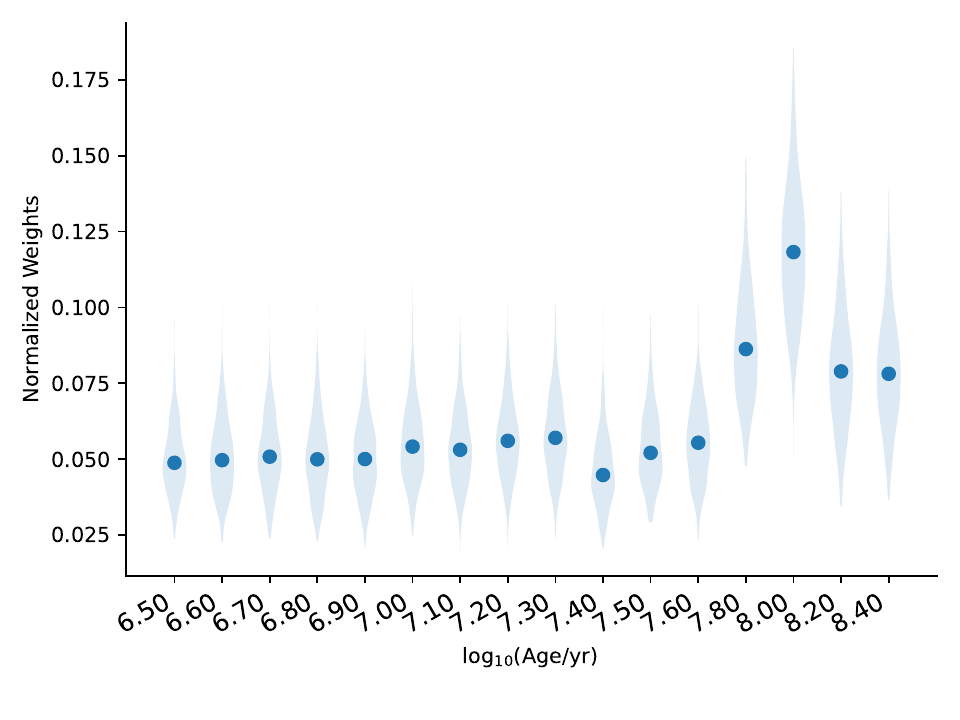}
    \end{minipage}
    \hfill
    \begin{minipage}[b]{0.32\textwidth}
        \centering
        \includegraphics[width=\textwidth]{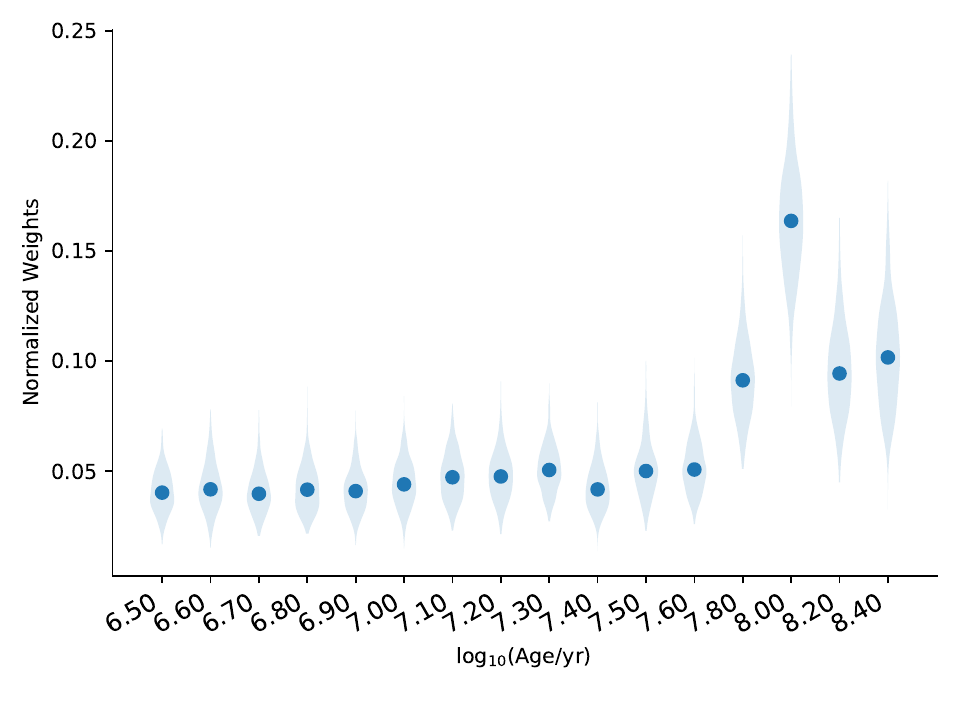}
    \end{minipage}
    \par\vspace{-1.02em}
    \caption{\small Inferred age weights for the datasets shown in Figure \ref{fig:CMDMostLikely}, marginalized over metallicity and extinction. The marginalized values of  $\tilde{A}_{\text{V}}$ range from 0.0 to 1.5, and $[M/H]$ range from -0.40 to 0.20. The panels from left to right show the 50pc, 100pc, and 150pc datasets respectively. Note that in the larger dataset (150 pc), the older population $\log_{10}$(t/yr) = 8.00 becomes more pronounced. At the same time, there is minor signal for a younger population with weights around $\log_{10}(t/{\rm yr}) = 7.30$.}
    \label{fig:AgeViolin}
\end{figure}

Figure~\ref{fig:AgeViolin} displays violin plots of marginalized age weights for each dataset, illustrating how inferred ages vary with circle radii. A significant signal for an older population emerges around $\log_{10}$(t/yr) = 8.00, becoming more pronounced with increasing radius. A smaller peak at $\log_{10}$(t/yr) = 7.30 reflects the ages of the few brightest evolved stars in Figure \ref{fig:CMDMostLikely}. Due to the age uncertainty for the rest of the stars in the field, one might mistakenly infer from Figure \ref{fig:AgeViolin} that the most likely age of the SN progenitor is $\log_{10}$(t/yr) = 8.00. However, these marginalized weights represent the entire population's age distribution. Since the SN progenitor likely resembled one of the brightest evolved stars, carefully examining Figure \ref{fig:CMDMostLikely} is essential to discern its true age.

\begin{figure}[H]
    \centering
    \includegraphics[width=\textwidth]{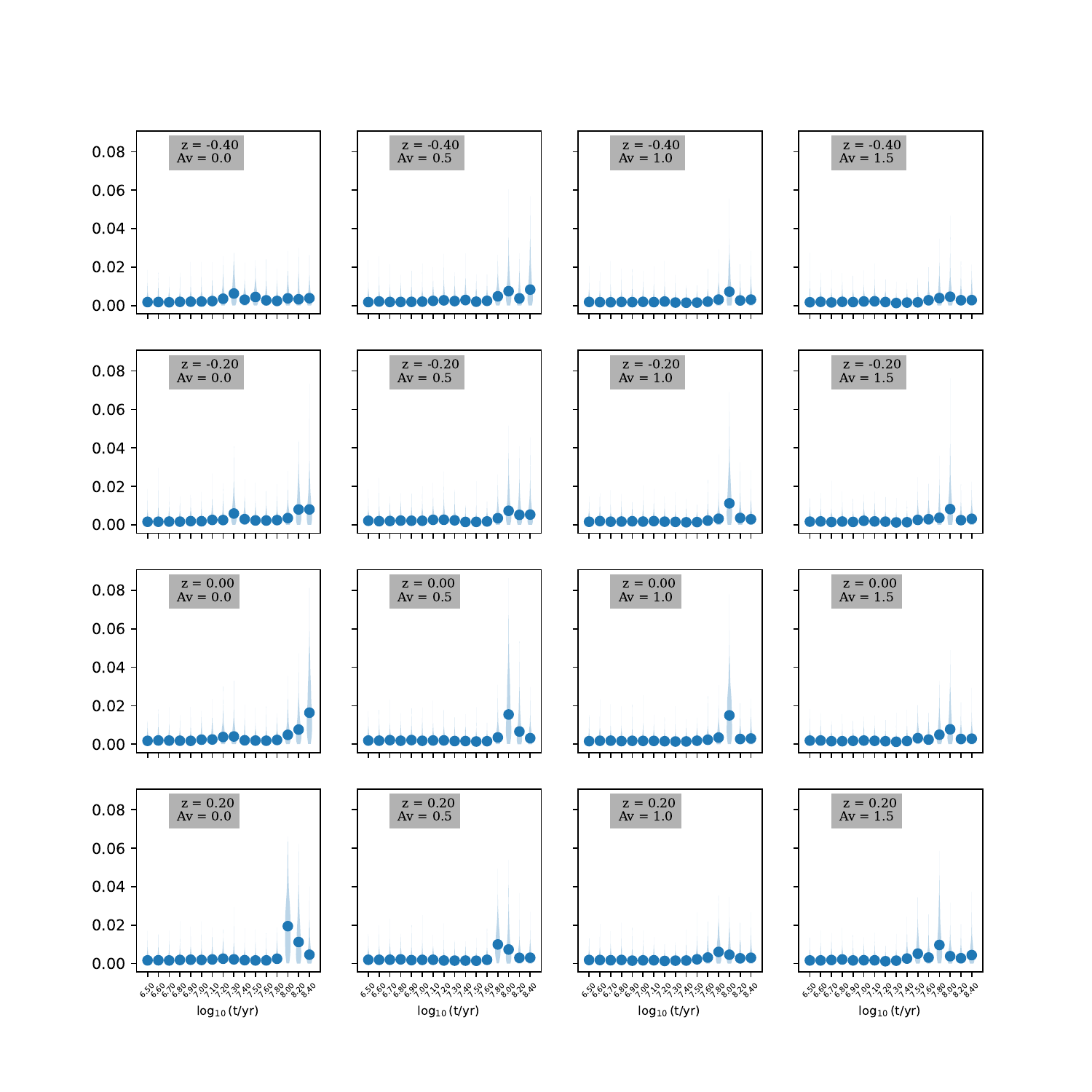}
    \par\vspace{-3.02em}
    \caption{Inferred weights as a function of age, metallicity, and extinction. This grid represents the un-marginalized inference for the 150pc dataset. From left to right, the columns present $A_{\text{V}}$ values of 0.0, 0.5, 1.0 and 1.5. From top to bottom, the rows present metallicity values of -0.4, -0.2, 0.0, and 0.2.}
    \label{fig:WeightsGrid}
\end{figure}

Figure \ref{fig:WeightsGrid} illustrates the posterior distribution for age, metallicity, and median extinction for the 150 pc dataset.
This grid visualizes how these parameters collectively influence our stellar age estimates. Across all cells, the $\log_{10}$(t/yr) $\sim$ 8.0 age signal is present, indicating a degeneracy for metallicity and extinction values for this population. When the younger $\log_{10}$(t/yr) = 7.3 age signal appears, the grid becomes more discerning, indicating a preference for lower metallicities ([M/H] ranging from -0.4 to 0.0) and low extinction around $\sim$ 0.0 (in other words, the top 3 cells in the left column of Figure \ref{fig:WeightsGrid}).

\begin{figure}[H]
    \centering
    \begin{minipage}[b]{0.32\textwidth}
        \centering
        \includegraphics[width=\textwidth]{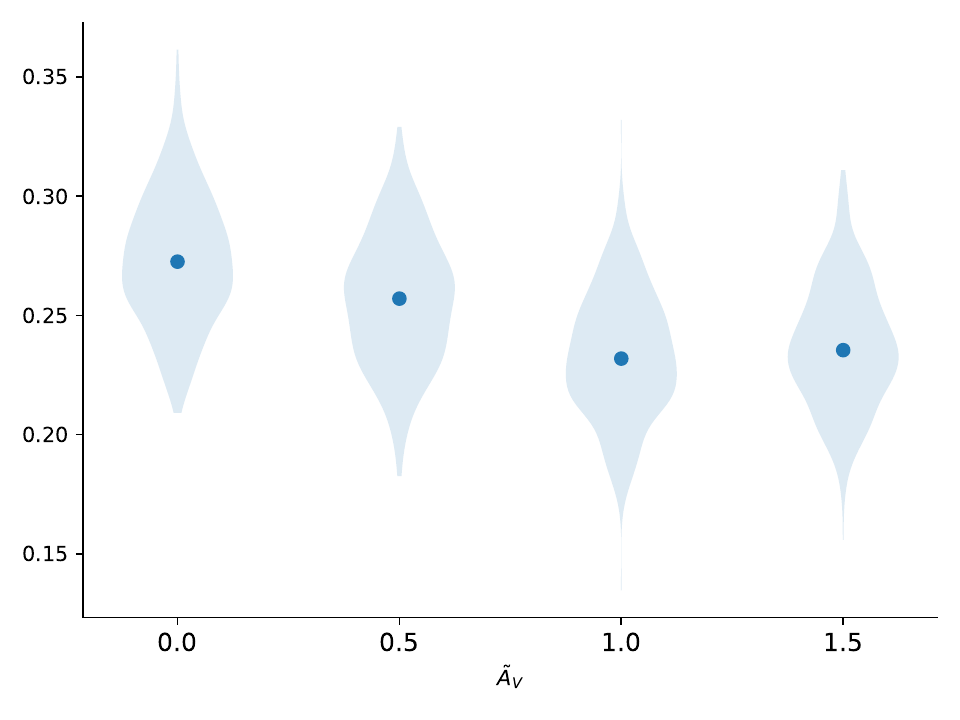}
    \end{minipage}
    \hfill
    \begin{minipage}[b]{0.32\textwidth}
        \centering
        \includegraphics[width=\textwidth]{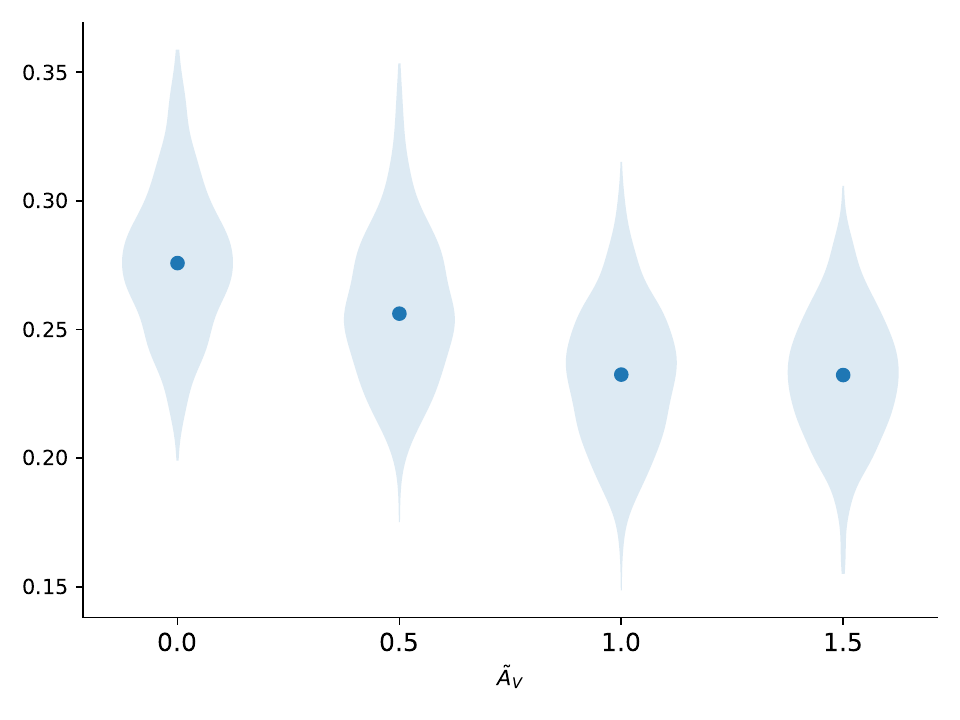}
    \end{minipage}
    \hfill
    \begin{minipage}[b]{0.32\textwidth}
        \centering
        \includegraphics[width=\textwidth]{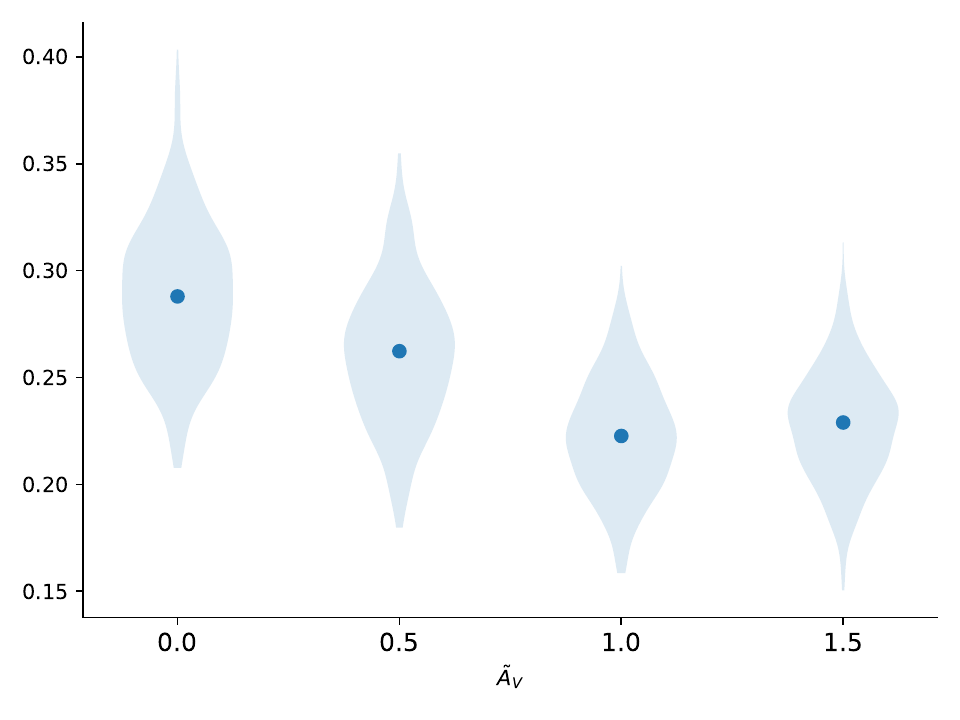}
    \end{minipage}
    \caption{\small Inferred extinction weights marginalized over age and metallicity. The marginalized ages range from log$_{10}$(t/yr) = $6.50$  to $8.40$, and [M/H] values range from -0.40 to 0.20. The panels from left to right show the 50pc, 100pc, and 150pc datasets respectively. For the 50 and 100 pc datasets, there is a slight preference for lower extinctions, while the larger dataset (150pc) shows a more pronounced preference for lower extinctions. \citet{Wang_2005} report an E(B-V) (mag) = $0.34 \pm 0.05$ for the surrounding star cluster. Our results, however, align more closely with \citet{Williams_2014} who suggest an $A_{\text{V}} = 0.11$}
    \label{fig:AvViolin}
\end{figure}

Figure \ref{fig:AvViolin} shows the marginalized extinction weights. These plots show how extinction varies across the datasets, allowing us to assess the spatial dependence of the extinction inference. The results indicate a trend towards lower extinctions as the circle radius increases. However, the large uncertainty in the weights suggests that there may not be sufficient information to constrain extinction for this population.

\begin{figure}[H]
    \centering
    \begin{minipage}[b]{0.45\textwidth}
        \centering
        \includegraphics[width=\textwidth]{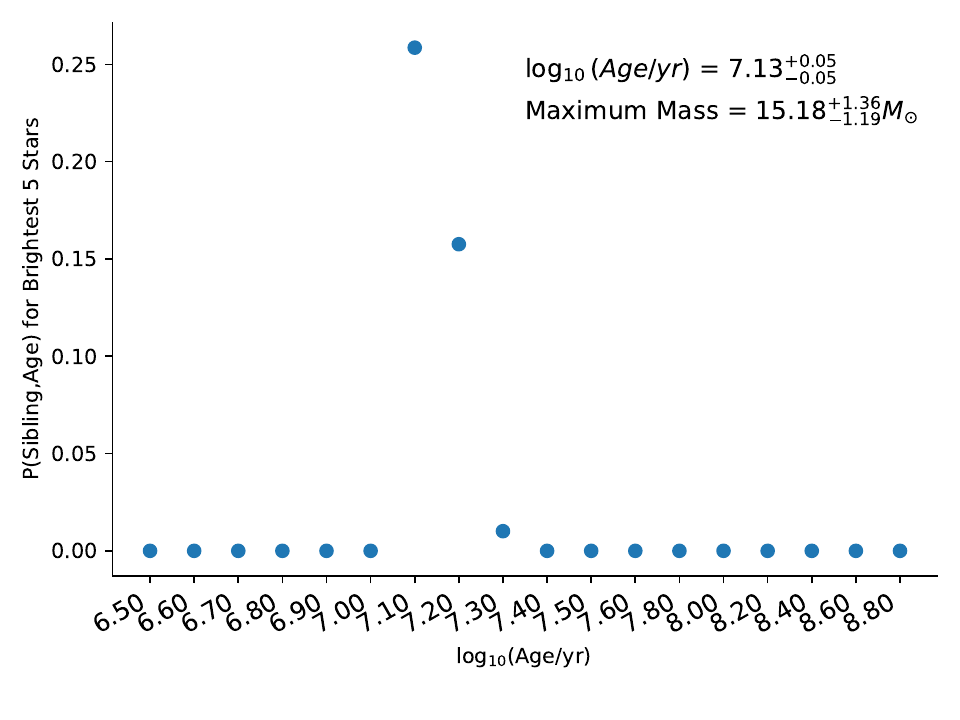}
    \end{minipage}
    \hfill
    \begin{minipage}[b]{0.45\textwidth}
        \centering
        \includegraphics[width=\textwidth]{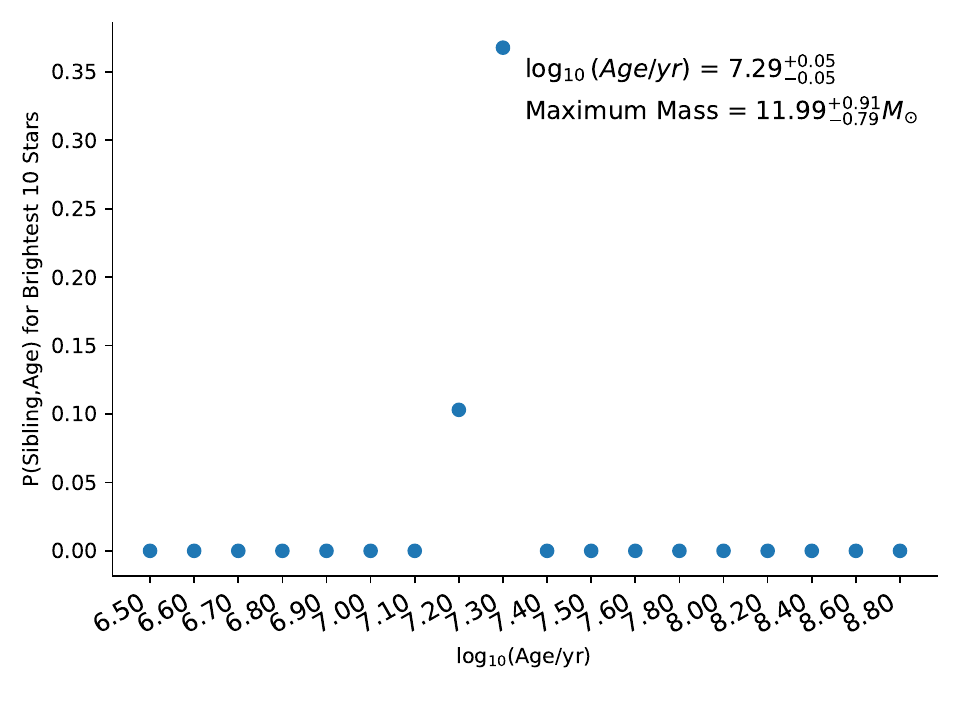}
    \end{minipage}
    \par\vspace{-1.02em}
    \caption{\small Inferred age weights and corresponding maximum mass utilizing only the brightest stars in the image, excluding extinction. The left panel uses the 5 brightest stars only, whereas the right panel uses the 10 brightest. \citet{Wang_2005} reported a turnoff mass of $\sim$ 12 $M_{\odot}$ using precursor imaging. \citet{Williams_2014} finds a median mass of 12.9 $M_{\odot}$ utilizing MATCH. Compared to Figure \ref{fig:AgeViolin}, the age (and therefore mass) inferences are more precise.}
    \label{fig:10BrightestTZ}
\end{figure}

\begin{figure}[H]
    \centering
    \begin{minipage}[b]{0.45\textwidth}
        \centering
        \includegraphics[width=\textwidth]{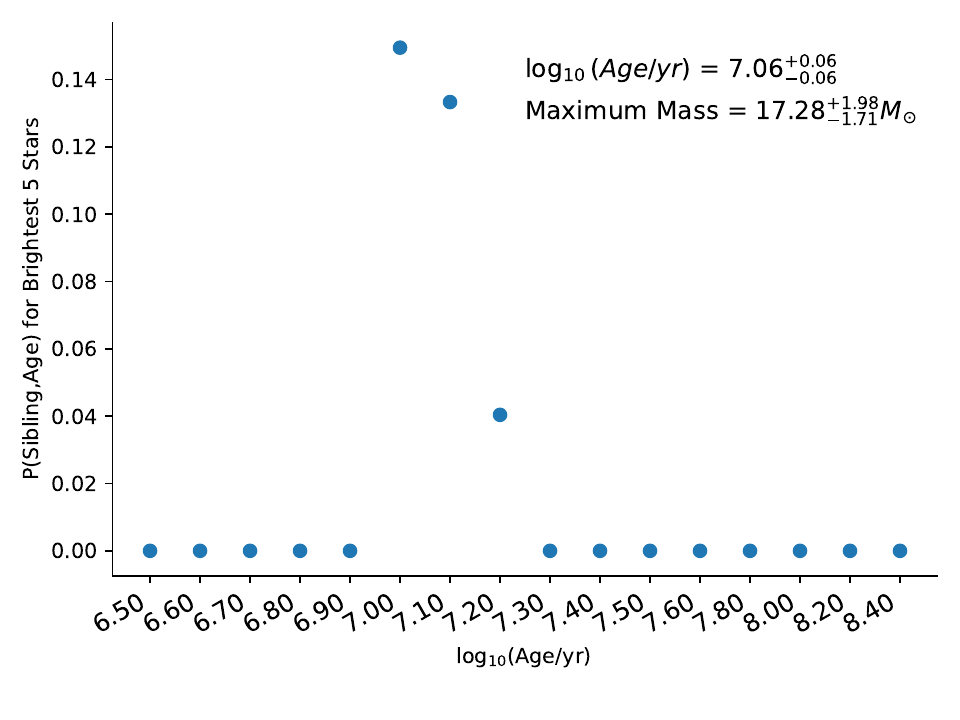}
    \end{minipage}
    \hfill
    \begin{minipage}[b]{0.45\textwidth}
        \centering
        \includegraphics[width=\textwidth]{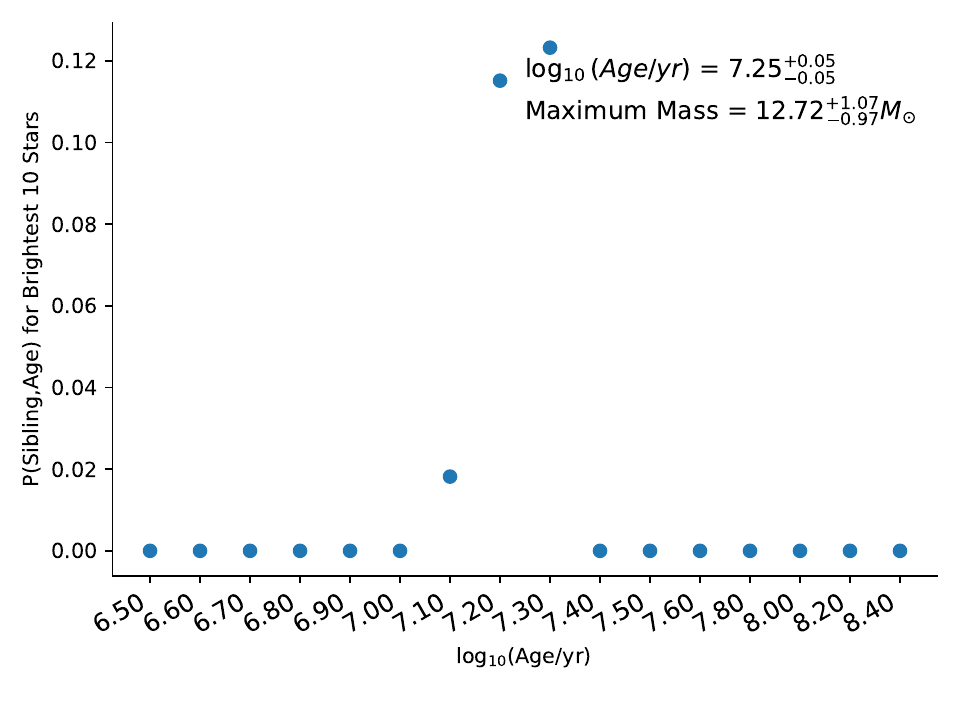}
    \end{minipage}
    \caption{\small Inferred age weights and corresponding mass using the brightest stars in the image, including extinction. Compared to Figure \ref{fig:10BrightestTZ}, including extinction increases the inferred maximum mass, as well as the variation in the inference between using the 5 or 10 brightest stars.}
    \label{fig:10BrightestTZA}
\end{figure}

The brightest evolved stars show a consistent age. By inferring each star's parameters individually using {\it Stellar Ages}, we can select these stars as potential siblings of the exploded star, leading to a more precise determination. Figures \ref{fig:10BrightestTZ} and \ref{fig:10BrightestTZA} present inferred ages using the brightest stars, both with and without extinction. Determining a sibling to the SN progenitor requires careful consideration, so we infer the age and maximum mass from a range of the brightest stars. Figure \ref{fig:10BrightestTZ} (left) shows results for the five brightest stars, while the right panel uses the ten brightest. In Figure \ref{fig:10BrightestTZA}, extinction is included, which increases the inferred maximum mass and highlights the impact of including the extra stars.

Using the median results from Figures \ref{fig:10BrightestTZ} and \ref{fig:10BrightestTZA}, we derive an age of $\log_{10}$(Age/yr) = $7.19^{+0.10}_{-0.13}$ and maximum mass of $13.95^{+3.33}_{-1.96}$ $\text{M}_{\odot}$. The ages span from $\log_{10}$(Age/yr) = $7.06$ to $7.29$, and the mass estimates are from $11.99 M_{\odot}$ to $17.28$ $\text{M}_{\odot}$, reflected in the reported uncertainties.
These results are all consistent with each other, showing that the conclusions remain robust across circle radii and extinction considerations. These results also roughly align with \citet{Williams_2014}, who found a median mass of 12.9 $\text{M}_{\odot}$ using MATCH, and \citet{Wang_2005} who reported a turnoff mass of approximately 12 $\text{M}_{\odot}$ from precursor imaging. Recently, \citet{beasor2024}, suggested that inferred progenitor masses from precursor imaging are systematically underestimated. Therefore, if we can generalize their result to our specific case of SN 2004dj, it would lend support for our slightly larger mass estimate. Thus, Figures \ref{fig:10BrightestTZ} and \ref{fig:10BrightestTZA} show that focusing on the brightest stars refines age and mass estimates, enhancing precision while maintaining consistency with previous findings.

\section{Discussion} \label{sec:discussion}

The results from our analysis underscore the utility and robustness of \textit{Stellar Ages} inferences. By applying it to both synthetic and real datasets, we have demonstrated that the technique produces reliable estimates of stellar ages and masses across diverse contexts. Our results highlight how focusing on the brightest stars enhances the precision of age and mass estimates, reducing uncertainties and are well-aligned with previous studies. This confirms that targeting such stars is a powerful approach for refining stellar population inferences, particularly when spatial parameters are factored into the analysis.

The case of SN 2004dj is a clear example of how \textit{Stellar Ages} can be used to infer progenitor characteristics, contributing to the broader understanding of supernova mechanisms and stellar evolution. By obtaining detailed information about the environment surrounding the supernova, we can draw meaningful conclusions about the final stages of stellar life and the formation of remnants. Our analysis has shown that the median age and mass estimates for SN 2004dj are in good agreement with previous work, strengthening the credibility of this method for such studies.

However, the accuracy of these estimates is inherently linked to the underlying models, such as PARSEC, MIST, and Geneva \citep{Lejeune2001_Geneva, Eggenberger2021_GenevaRot}, each with its own set of assumptions and limitations. Therefore, the reliability of \textit{Stellar Ages} is contingent on the models it employs. Moreover, our multi-parameter model introduces uncertainties that may reflect potential over fitting. Fixing parameters like metallicity or extinction, when reliable priors are available, could mitigate these issues and enhance the robustness of our analysis.

Rotation models represent another challenge, as current models (e.g. MIST, PARSEC, Geneva) only accommodate a single rotation value rather than a distribution \citep{Wright_2011_rotation, Sun2024_rotation}. This simplification may not adequately capture the full complexity of stellar populations, limiting the model's precision in certain cases.

Marginalizing weights across the entire stellar population can obscure specific subsets of interest. For example, when inferring the age of stars that exploded, or other late-stage evolution products, there is a high probability that the brightest, evolved stars are siblings of the star in question. Therefore, using the brightest, evolved stars can significantly reduce age variation, improving the reliability of inferred properties. Incorporating greater refinement of sibling determination into future applications of \textit{Stellar Ages} could yield even more precise results, particularly in studies where these stellar subsets are critical.

In summary, while \textit{Stellar Ages} has demonstrated its utility, further refinements—such as handling rotation more effectively and reducing potential over fitting, will improve its accuracy. These refinements will make the code a versatile tool for investigating stellar evolution, supernova progenitors, and the broader structure of galaxies.

\section{Conclusion}
\label{sec:Conclusion}

We introduced and validated the \textit{Stellar Ages} algorithm, demonstrating its ability to accurately infer stellar ages, metallicities, and extinctions using joint probability density functions and Gibbs MCMC sampling. Tests on both synthetic and real datasets, including the stellar population surrounding SN 2004dj, confirmed its reliability in inferring stellar ages and progenitor masses. The results were consistent with previous studies, demonstrating that \textit{Stellar Ages} may yield more precise age determinations than traditional methods, particularly when focusing on the brightest stars.

The development of \textit{Stellar Ages} marks a step forward in stellar population analysis, offering broad applicability across different astronomical datasets and observatories. By enhancing the accuracy and precision of age estimates, this tool holds potential for advancing our understanding of stellar evolution and the life cycles of stars. Moreover, the unique capability of \textit{Stellar Ages} to handle individual stars and the broader population simultaneously opens new possibilities for detailed investigations into stellar formation and death.

\textit{Stellar Ages} shifts the statistical question from ``What is the expected number of stars within the CMD?'' to ``What is the most likely age of each star?'' This shift enables more effective utilization of all available information for a small set of stars.  For example, the presence of a small set of evolved stars strongly implies a specific age for that group.  If the progenitor is associated with those stars, then we can produce more precise statistical inferences.

In conclusion, \textit{Stellar Ages} not only refines existing techniques but also introduces new methodologies for the astronomical community to explore the complexities of stellar populations and their evolutionary paths.

\section*{Acknowledgments}
This work was supported by NASA through grant HST-GO-16778.017-A from the Space Telescope Science Institute, which is operated by the Association of Universities for Research in Astronomy, Inc., under NASA contract NAS 5-26555.

Some/all of the data presented in this paper were obtained from the Mikulski Archive for Space Telescopes (MAST) at the Space Telescope Science Institute. The specific observations analyzed can be accessed via \dataset[https://doi.org/10.17909/hae1-qj98]{https://doi.org/10.17909/hae1-qj98}. STScI is operated by the Association of Universities for Research in Astronomy, Inc., under NASA contract NAS5–26555. Support to MAST for these data is provided by the NASA Office of Space Science via grant NAG5–7584 and by other grants and contracts.

\vspace{5mm}
\facilities{HST(ACS), MAST}

\setlength{\bibsep}{0pt}
\bibliographystyle{aasjournal}
\bibliography{joe}{}

\end{document}